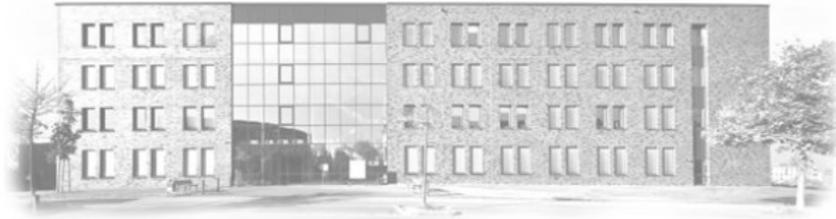

TECHNICAL REPORTS IN COMPUTER SCIENCE
Technische Universität Dortmund

# Generalized permutations and Ternary Bent Functions

**Claudio Moraga**

Lehrstuhl Informatik I
Logik in der Informatik
University of Dortmund
Germany

Claudio.Moraga@udo.edu

**Milena Stanković**

Faculty of
Electronic Engineering
University of Niš
Serbia

Milena.Stankovic@elfak.ni.ac.rs

**Radomir Stanković**

Institute of Mathematics
Serbian Academy
of Sciences and Arts
Serbia

Radomir.Stankovic@gmail.com

# Generalized Permutations and Ternary Bent Functions

Abstract: The report studies the generation of ternary bent functions by permuting the circular Vilenkin-Chrestenson spectrum of a known bent function. We call this *spectral invariant operations in the spectral domain*, in analogy to the spectral invariant operations in the domain of the functions. Furthermore, related generalized permutations are derived to obtain new bent functions in the original domain. In the case of 2-place ternary bent functions a class of permutations with a Kronecker product structure is disclosed, which allows generating all 2-place ternary bent functions, based on a set of 9 "seed" functions.

## 1 INTRODUCTION

Bent functions [1, 2, 7, 11, 13, 15, 18, 19] are functions with maximal non-linearity, of particular interest for cryptography, combinatorics and coding theory. From the mathematical point of view they offer several challenging problems, like characterization, generation, classification and counting, and generation of spectral invariant operations. This last one being the subject of this report.

Spectral Invariant Operations constitute a set of operations, which when applied to bent functions produce new bent functions. Their effect observed in the spectral domain is, that they preserve the flatness of the corresponding spectra. Flatness consists of the constant magnitude of the spectral components, which is a characterizing property of bent functions. In the *p*-valued case, the magnitude of the spectral components is $p^{n/2}$. In other words these invariant operations induce a generalized permutation[1] of the spectra of the processed functions. In the binary case, bent functions are characterized by the Walsh-Hadamard spectrum, whereas in the *p*-valued case they are characterized by the Vilenkin-Chrestenson [20], [3] spectrum [11].

Looking at the past, Invariance Theorems in Switching Theory may be found in [8] and the first five Spectral Invariant Operations for the binary case may be traced back to the work of S.L. Hurst [5]. This was soon extended to the ternary domain in [9, 10] and to multiple-valued functions in [16]. A recent review and further extension may be found in [18]. In all these *p*-valued cases care was taken to identify the corresponding spectral coefficients that are affected after a spectral invariant operation is applied to a given function. It was shown that the corresponding spectrum was affected by a generalized permutation, but no explicit permutation matrix was given (which, as will be shown, may not necessarily be unique). In this report the opposite situation is considered. Given the characterizing spectrum of a bent function, does a generalized permutation of this spectrum lead to a new bent function? If yes, the permutation constitutes a spectral invariant operation in the spectral domain.

---

[1] A square generalized permutation matrix has a single non-zero element in each row and column. The non-zero entries may be 1 or a power of $\xi$, a primitive *p*-th root of unity. All entries may be scaled by $\pm 1$.

## 2 FORMALISMS

Let A denote a square matrix. The notation A($n$) will be used to represent a $p^n \times p^n$ matrix named A.

Definition 1:
The basic Vilenkin-Chrestenson transform matrix is called C(1) = [$c_{j,k}$], with $j, k \in \mathbb{Z}_p$ and the value of $c_{j,k}$ is obtained as $\xi^{j \cdot k \bmod p}$, where $\xi$ denotes a primitive $p$-th root of unity.

For $p$ = 3, C(1) is defined as follows:

$$C(1) = \begin{bmatrix} 1 & 1 & 1 \\ 1 & \xi & \xi^2 \\ 1 & \xi^2 & \xi \end{bmatrix} \tag{1a}$$

where $\xi$ denotes a primitive 3-*rd* root of unity. It is simple to show that $\xi = \cos(2\pi/3) + i \sin(2\pi/3) = -0.5 + i\sqrt{3}/2$, where $i = \sqrt{-1}$. Moreover $\xi^2$ equals $\xi^*$, the complex conjugate of $\xi$ [6], [17].

For $p$ = 4, C(1) is defined as follows:

$$C(1) = \begin{bmatrix} 1 & 1 & 1 & 1 \\ 1 & \xi & \xi^2 & \xi^3 \\ 1 & \xi^2 & 1 & \xi^2 \\ 1 & \xi^3 & \xi^2 & \xi \end{bmatrix} \tag{1b}$$

In this case $\xi = \cos(\pi/2) + i \sin(\pi/2)$, which obviously leads to $\xi = i$, $\xi^2 = -1$ and $\xi^3 = -i$.

Furthermore, for all $p$, the Vilenkin-Chrestenson transform matrix has a Kronecker product structure:

$$C(n) = C(n\text{-}1) \otimes C(1) = C(1) \otimes C(n\text{-}1) = C(1)^{\otimes n} \tag{2}$$

where $C(1)^{\otimes n}$ denotes the $n$-fold Kronecker product of C(1) with itself. Finally, C($n$) is orthogonal: $C(n) \cdot C^*(n) = p^n I(n)$ [6], where I($n$) is the $p^n \times p^n$ identity matrix.

When no confusion arises, $f$ will denote the zero-polarity Reed-Muller polynomial expression of a ternary function or its value vector. Its sign representation will be denoted by $F$. (See Definition 2). In what follows, the needed analyses will be done with $p$ = 3, but the methods are extendable to $p > 3$, not necessarily a prime. Also, functions of two variables will be considered, however some of the presented results may be extended to larger number of variables. If $f$ is a ternary bent function its spectral characterization is given by the Vilenkin-Chrestenson spectrum of $\xi^{f(x)}$, called circular spectrum of $f$ in [11] or spectrum of the sign function in [1, 2]. All components of this spectrum have the same magnitude, equal to $3^{n/2}$. It is said that the spectrum is "flat". Flatness of the spectrum is a necessary condition for a function to be bent. It is however not sufficient, in the sense that given a vector of $p^n$ entries, all of them with magnitude $p^{n/2}$, the inverse transform of this vector (see Eqs. (5) and (7) below) does not necessarily return a bent function in a sign expression. The number of generalized permutations on a (flat) Vilenkin-Chrestenson spectrum is much larger than the number of bent functions for the corresponding number of variables. For $p$ = 3 and $n$ = 2, the number of only classical permutations of the coefficients of the spectra of these functions amounts to 9! = 362,880, whereas the number of ternary bent functions of 2 variables is 486 [11]. In what follows, classical permutations will be simply called straight.

Definition 2:

If $f$ is an $n$-place ternary bent function, $F$ will denote the value vector of its sign $\xi^{f(x)}$; i.e.

$$F = [\xi^{f(0)}, \xi^{f(1)}, \ldots, \xi^{f(3^n-2)}, \xi^{f(3^n-1)}]^T \quad (3)$$

Definition 3: See, e.g. [11].

Let $f$ be an $n$-place ternary function. Its circular Vilenkin-Chrestenson spectrum (or the Vilenkin-Chrestenson spectrum of its sign function $\xi^{f(x)}$) as well as its inverse are given by:

$$S_f(w) = \sum_{x=0}^{3^n-1} \xi^{f(x) - \langle w \cdot x \rangle} \quad (4)$$

and

$$\xi^{f(x)} = 3^{-n} \sum_{w=0}^{3^n-1} S_f(w) \cdot \xi^{\langle w \cdot x \rangle}, \quad (5)$$

where $\langle w \cdot x \rangle$ denotes the scalar product of $w$ and $x$.

In matrix notation, with C*($n$) denoting the complex conjugate of C($n$),

$$S_f = C^*(n) \cdot F \quad (6)$$

$$F = 3^{-n} C(n) \cdot S_f. \quad (7)$$

## 3 ANALYSIS OF CASES

Instead of searching for new spectral invariant operations (in the domain of bent functions) in this report we are considering the other side of the coin: We are interested in the following type of question: Let $f$ be an $n$-place ternary bent function with spectrum $S_f$. Does there exist a ternary bent function $g$ such that $S_g = P(n) \cdot S_f$, where P($n$) is a generalized permutation matrix? If yes, does the new function exhibit particular structural features? A straight forward method to solve the first part of this kind of problems is to check whether the candidate spectrum is flat and apply the inverse transform to it. If the result is the value vector of the sign of a ternary bent function, the answer will be yes. This method may be considered to be efficient; but it does not give any further information. Instead we will consider a step by step analysis.

The general case corresponds to finding a ternary bent function $g$, where $S_g$ denotes the Vilenkin-Chrestenson spectrum of its sign representation, such that $S_g = P(n) S_f$. In this expression P($n$) is a generalized permutation matrix and $S_f$ denotes the spectrum of the sign representation of a reference bent function $f$.

From (7), $G = 3^{-n} C(n) \cdot S_g$ and from the given condition,

$$G = 3^{-n} C(n) \cdot P(n) \cdot S_f. \quad (8)$$

Let

$$W(n) = 3^{-n} C(n) \cdot P(n) \cdot C^*(n), \quad (9)$$

from where
$$W(n) \cdot C(n) = C(n) \cdot P(n). \tag{10}$$

Introducing (10) in (8)
$$G = 3^{-n} W(n) C(n) \cdot S_f = W(n) \cdot ( 3^{-n} C(n) \cdot S_f ) = W(n) \cdot F. \tag{11}$$

**Case 1**:

Let $f$ be an $n$-place ternary bent function with spectrum $S_f$. Does there exist a ternary bent function $g$ such that $S_g = (S_f)^*$ ?

To solve the problem with the general approach introduced above, a generalized permutation should be found, such that $P(n) S_f = (S_f)^*$. Recalling that $\xi^2$ equals $\xi^*$, and $(\xi^2)^* = (\xi^*)^* = \xi$, it may be concluded that $P(n) = 3^{-n/2} \, diag(S_f)$. As will be shown later, depending on the structure of $S_f$, other generalized permutation matrices may be found.

Then, according to (9)
$$W(n) = 3^{-n} C(n) \cdot ( 3^{-n/2} \, diag(S_f)) \cdot C^*(n),$$

and with (11),
$$G = 3^{-n} C(n) \cdot ( 3^{-n/2} \, diag(S_f)) \cdot C^*(n) \cdot F = W(n) \cdot F \tag{12}$$

For the posed problem the following formal expression may also be used.
$$G = 3^{-n} C(n) \cdot S_g = 3^{-n} C(n) \cdot (S_f)^* = 3^{-n} (C(n)^* \cdot S_f)^* \tag{13}$$

Notice that $C^*(1) = \begin{bmatrix} 1 & 1 & 1 \\ 1 & \xi^2 & \xi \\ 1 & \xi & \xi^2 \end{bmatrix}$ and let $P_{12}(1) = \begin{bmatrix} 1 & 0 & 0 \\ 0 & 0 & 1 \\ 0 & 1 & 0 \end{bmatrix} = P_{12}^*(1)$, with $P_{12}(n) = (P_{12}(1))^{\otimes n}$.

Then
$$C^*(1) = P_{12}(1) \cdot C(1) \text{ and } C^*(n) = P_{12}(n) \cdot C(n). \tag{14}$$

Introducing (14) in (13):
$$G = 3^{-n} \langle P_{12}(n) \cdot C(n) \cdot S_f \rangle^* = P_{12}(n) \cdot 3^{-n} \langle C(n) \cdot S_f \rangle^* = P_{12}(n) \cdot F^*. \tag{15}$$

Eqs. (12) and (15) provide a different kind of information on $G$ than the calculation of the inverse Vilenkin-Chrestenson spectrum of $(S_f)^*$.

**Example 1**: For the posed problem, let $f = x_1 x_2$, which is the simplest 2-place ternary bent function. A direct calculation gives:

$S_f = [3 \; 3 \; 3 \; 3 \; 3\xi^2 \; 3\xi \; 3 \; 3\xi \; 3\xi^2 \;]^T = 3[\,1 \; 1 \; 1 \; 1 \; \xi^2 \; \xi \; 1 \; \xi \; \xi^2 \,]^T$

from where $P(2) = diag(1 \; 1 \; 1 \; 1 \; \xi^2 \; \xi \; 1 \; \xi \; \xi^2\,)$

Recalling the Pauli matrices [14],

$Z(1) = \begin{bmatrix} 1 & 0 & 0 \\ 0 & \xi & 0 \\ 0 & 0 & \xi^2 \end{bmatrix}$, with $Z(n) = Z(1)^{\otimes n}$ and $X(1) = \begin{bmatrix} 0 & 0 & 1 \\ 1 & 0 & 0 \\ 0 & 1 & 0 \end{bmatrix}$, with $X(n) = X(1)^{\otimes n}$

it is simple to see that the matrix P(2) may also be specified as

P(2) = *blockdiag* ( I(1) , Z*(1) , Z(1) ).

Different specifications do not change the corresponding matrix, but may affect the complexity of calculating W(*n*). (See equations (33) to (45)).

Calculating *G* with eq. (12) leads to the following

$$
\overset{W(2)}{\begin{bmatrix}
1 & 1 & 1 & 1 & \xi & \xi^2 & 1 & \xi^2 & \xi \\
1 & 1 & 1 & \xi^2 & 1 & \xi & \xi & 1 & \xi^2 \\
1 & 1 & 1 & \xi & \xi^2 & 1 & \xi^2 & \xi & 1 \\
1 & \xi^2 & \xi & 1 & 1 & 1 & 1 & \xi & \xi^2 \\
\xi & 1 & \xi^2 & 1 & 1 & 1 & \xi^2 & 1 & \xi \\
\xi^2 & \xi & 1 & 1 & 1 & 1 & \xi & \xi^2 & 1 \\
1 & \xi & \xi^2 & 1 & \xi^2 & \xi & 1 & 1 & 1 \\
\xi^2 & 1 & \xi & \xi & 1 & \xi^2 & 1 & 1 & 1 \\
\xi & \xi^2 & 1 & \xi^2 & \xi & 1 & 1 & 1 & 1
\end{bmatrix}} \cdot \overset{F}{\begin{bmatrix} 1 \\ 1 \\ 1 \\ 1 \\ \xi \\ \xi^2 \\ 1 \\ \xi^2 \\ \xi \end{bmatrix}} = \overset{G}{\begin{bmatrix} 1 \\ 1 \\ 1 \\ 1 \\ \xi^2 \\ \xi \\ 1 \\ \xi \\ \xi^2 \end{bmatrix}} \quad (16)
$$

The steps to calculate *G* with eq. (15) are summarized in Table 1.

TABLE 1: STEPS LEADING FROM $f$ TO $g$.

| $f$ | $F$ | $S_f$ | $S_g = S_f^*$ | $P_{12}(2) \cdot F^*$ | $g$ |
|---|---|---|---|---|---|
| 0 | 1 | 3 | 3 | 1 | 0 |
| 0 | 1 | 3 | 3 | 1 | 0 |
| 0 | 1 | 3 | 3 | 1 | 0 |
| 0 | 1 | 3 | 3 | 1 | 0 |
| 1 | $\xi$ | $3\xi^2$ | $3\xi$ | $\xi^2$ | 2 |
| 2 | $\xi^2$ | $3\xi$ | $3\xi^2$ | $\xi$ | 1 |
| 0 | 1 | 3 | 3 | 1 | 0 |
| 2 | $\xi^2$ | $3\xi$ | $3\xi^2$ | $\xi$ | 1 |
| 1 | $\xi$ | $3\xi^2$ | $3\xi$ | $\xi^2$ | 2 |

It is simple to see that $g = 2x_1x_2$, which is also bent. Notice that its spectrum is flat and the components have magnitude 3, which corresponds to $3^{n/2}$, since $n = 2$.

In this case, the spectral analysis leads to a new function that may be given different formal expressions based on a single spectral invariant operation in the domain of the function. It also gives two explicit permutations –($P_{12}(n)$ or $W(n)$)– leading to the solution

$g = 2(x_1x_2) = (2\ x_1)\ x_2 = x_1\ (2\ x_2) = 2(2\ x_1)(2\ x_2)$.

Notice that $P_{12}(n)$ is not the generalized permutation matrix, which initiated the problem.

**Case 2**: Let $N(1) = \begin{bmatrix} 0 & 0 & 1 \\ 0 & 1 & 0 \\ 1 & 0 & 0 \end{bmatrix}$ and $N(n) = \langle N(1) \rangle^{\otimes n}$

Let $f$ be an $n$-place ternary bent function with spectrum $S_f$. Does there exist a ternary bent function $g$ such that $S_g = N(n) S_f$ ?

$G = 3^{-n} C(n) \cdot S_g = 3^{-n} C(n) \cdot N(n) \cdot S_f = 3^n \langle C(1)N(1) \rangle^{\otimes n} \cdot S_f$.

Notice that

$$C(1)N(1) = \begin{bmatrix} 1 & 1 & 1 \\ 1 & \xi & \xi^2 \\ 1 & \xi^2 & \xi \end{bmatrix} \cdot \begin{bmatrix} 0 & 0 & 1 \\ 0 & 1 & 0 \\ 1 & 0 & 0 \end{bmatrix} = \begin{bmatrix} 1 & 1 & 1 \\ \xi^2 & \xi & 1 \\ \xi & \xi^2 & 1 \end{bmatrix}$$

and (17)

$$3^{-1}C(1)N(1)C^*(1) = 3^{-1} \begin{bmatrix} 1 & 1 & 1 \\ \xi^2 & \xi & 1 \\ \xi & \xi^2 & 1 \end{bmatrix} \cdot \begin{bmatrix} 1 & 1 & 1 \\ 1 & \xi^2 & \xi \\ 1 & \xi & \xi^2 \end{bmatrix} = \begin{bmatrix} 1 & 0 & 0 \\ 0 & 0 & \xi^2 \\ 0 & \xi & 0 \end{bmatrix}$$

Recall that, since P(n) in this case equals N(n),

$$W(n) = 3^{-n} C(n)N(n)C^*(n) \tag{18}$$

$$W(n)C(n) = 3^{-n} C(n)N(n)C^*(n)C(n) = C(n)N(n) \tag{19}$$

Introducing (19) in the original equation for $G$ leads to:

$$G = 3^{-n} C(n) \cdot N(n) \cdot S_f = 3^{-n} W(n)C(n)S_f = W(n)(3^{-n} C(n)S_f) = W(n) \cdot F \tag{20}$$

From (9), (19) and the Kronecker product structure of the matrices, the following is obtained:

$$W(n) = 3^{-n} C(n)N(n)C^*(n) = \langle 3^{-1}C(1)N(1)C^*(1) \rangle^{\otimes n} = \begin{bmatrix} 1 & 0 & 0 \\ 0 & 0 & \xi^2 \\ 0 & \xi & 0 \end{bmatrix}^{\otimes n}. \tag{21}$$

Noltice that in this case W($n$) is a generalized permutation matrix.

**Example 2**: Consider again $f = x_1 x_2$, which is bent.

Table 2 summarizes the calculations to obtain $g$.

It is simple to see that $g = [\,0\ 2\ 1\ 2\ 2\ 2\ 1\ 2\ 0\,]^T = x_1 x_2 \oplus 2 x_1 \oplus 2 x_2$. This is also a bent function, since the spectrum of $\xi^{g(x)}$ is flat and the components have magnitude 3. Notice that working in the domain of the function, $g$ it is obtained applying 2 spectral invariant operations, whereas in the spectral domain, only one permutation was required.

TABLE 2: STEPS LEADING FROM $f$ TO $g$ IN CASE 2

| $f$ | $F$ | $S_f$ | $S_g$ | $W(n) \cdot F$ | $g$ |
|---|---|---|---|---|---|
| 0 | 1 | 3 | $3\xi^2$ | 1 | 0 |
| 0 | 1 | 3 | $3\xi$ | $\xi^2$ | 2 |
| 0 | 1 | 3 | 3 | $\xi$ | 1 |
| 0 | 1 | 3 | $3\xi$ | $\xi^2$ | 2 |
| 1 | $\xi$ | $3\xi^2$ | $3\xi^2$ | $\xi^2$ | 2 |
| 2 | $\xi^2$ | $3\xi$ | 3 | $\xi^2$ | 2 |
| 0 | 1 | 3 | 3 | $\xi$ | 1 |
| 2 | $\xi^2$ | $3\xi$ | 3 | $\xi^2$ | 2 |
| 1 | $\xi$ | $3\xi^2$ | 3 | 1 | 0 |

**Case 3**:

Let P(2) be a generalized permutation diagonal matrix:

$$P(2) = diag(\,\xi^2\ 1\ \xi\ 1\ 1\ 1\ \xi\ 1\ \xi^2\,). \tag{22}$$

Given $f = x_1x_2$ with spectrum $S_f$, does there exist a ternary bent function $g$ such that $S_g = P(2) \cdot S_f$? It is easy to see from (22) that $P(2) \cdot S_f$ is a flat spectrum and its components have magnitude 3.

$$G = 3^{-2} C(2) S_g = 3^{-2} C(2) P(2) S_f. \qquad (23)$$

Recall from (9)

$$W(2) = 3^{-2} C(2) P(2) C^*(2). \qquad (24)$$

Notice that

$$P(2) C^*(2) = D(2) \# C^*(2), \qquad (25)$$

where $D(2)$ is a column vector which elements are the diagonal elements of $P(2)$ and $\#$ denotes the extended Hadamard product[2] [4].

Then:

$$3^2 W(2) = C(2) P(2) C^*(2) =$$

$$= C(2) \cdot \left( \begin{bmatrix} \xi^2 \\ 1 \\ \xi \\ 1 \\ 1 \\ 1 \\ \xi \\ 1 \\ \xi^2 \end{bmatrix} \# \begin{bmatrix} 1 & 1 & 1 & 1 & 1 & 1 & 1 & 1 & 1 \\ 1 & \xi^2 & \xi & 1 & \xi^2 & \xi & 1 & \xi^2 & \xi \\ 1 & \xi & \xi^2 & 1 & \xi & \xi^2 & 1 & \xi & \xi^2 \\ 1 & 1 & 1 & \xi^2 & \xi^2 & \xi^2 & \xi & \xi & \xi \\ 1 & \xi^2 & \xi & \xi^2 & \xi & 1 & \xi & 1 & \xi^2 \\ 1 & \xi & \xi^2 & \xi^2 & 1 & \xi & \xi & \xi^2 & 1 \\ 1 & 1 & 1 & \xi & \xi & \xi & \xi^2 & \xi^2 & \xi^2 \\ 1 & \xi^2 & \xi & \xi & 1 & \xi^2 & \xi^2 & \xi & 1 \\ 1 & \xi & \xi^2 & \xi & \xi^2 & 1 & \xi^2 & 1 & \xi \end{bmatrix} \right) =$$

$$= C(2) \cdot \begin{bmatrix} \xi^2 & \xi^2 & \xi^2 & \xi^2 & \xi^2 & \xi^2 & \xi^2 & \xi^2 & \xi^2 \\ 1 & \xi^2 & \xi & 1 & \xi^2 & \xi & 1 & \xi^2 & \xi \\ \xi & \xi^2 & 1 & \xi & \xi^2 & 1 & \xi & \xi^2 & 1 \\ 1 & 1 & 1 & \xi^2 & \xi^2 & \xi^2 & \xi & \xi & \xi \\ 1 & \xi^2 & \xi & \xi^2 & \xi & 1 & \xi & 1 & \xi^2 \\ 1 & \xi & \xi^2 & \xi^2 & 1 & \xi & \xi & \xi^2 & 1 \\ \xi & \xi & \xi & \xi^2 & \xi^2 & \xi^2 & 1 & 1 & 1 \\ 1 & \xi^2 & \xi & \xi & 1 & \xi^2 & \xi^2 & \xi & 1 \\ \xi^2 & 1 & \xi & 1 & \xi & \xi^2 & \xi & \xi^2 & 1 \end{bmatrix} =$$

$$= \begin{bmatrix} 1 & 1 & 1 & 1 & 1 & 1 & 1 & 1 & 1 \\ 1 & \xi & \xi^2 & 1 & \xi & \xi^2 & 1 & \xi & \xi^2 \\ 1 & \xi^2 & \xi & 1 & \xi^2 & \xi & 1 & \xi^2 & \xi \\ 1 & 1 & 1 & \xi & \xi & \xi & \xi^2 & \xi^2 & \xi^2 \\ 1 & \xi & \xi^2 & \xi & \xi^2 & 1 & \xi^2 & \xi & 1 \\ 1 & \xi^2 & \xi & \xi & 1 & \xi^2 & \xi^2 & 1 & \xi \\ 1 & 1 & 1 & \xi^2 & \xi^2 & \xi^2 & \xi & \xi & \xi \\ 1 & \xi & \xi^2 & \xi^2 & \xi & 1 & \xi & \xi^2 & 1 \\ 1 & \xi^2 & \xi & \xi^2 & 1 & \xi & \xi & 1 & \xi^2 \end{bmatrix} \cdot \begin{bmatrix} \xi^2 & \xi^2 & \xi^2 & \xi^2 & \xi^2 & \xi^2 & \xi^2 & \xi^2 & \xi^2 \\ 1 & \xi^2 & \xi & 1 & \xi^2 & \xi & 1 & \xi^2 & \xi \\ \xi & \xi^2 & 1 & \xi & \xi^2 & 1 & \xi & \xi^2 & 1 \\ 1 & 1 & 1 & \xi^2 & \xi^2 & \xi^2 & \xi & \xi & \xi \\ 1 & \xi^2 & \xi & \xi^2 & \xi & 1 & \xi & 1 & \xi^2 \\ 1 & \xi & \xi^2 & \xi^2 & 1 & \xi & \xi & \xi^2 & 1 \\ \xi & \xi & \xi & \xi^2 & \xi^2 & \xi^2 & 1 & 1 & 1 \\ 1 & \xi^2 & \xi & \xi & 1 & \xi^2 & \xi^2 & \xi & 1 \\ \xi^2 & 1 & \xi & 1 & \xi & \xi^2 & \xi & \xi^2 & 1 \end{bmatrix} =$$

---

[2] In a generalized Hadamard product of a vector and a matrix, each element of the vector multiplies all elements of the corresponding matrix row.

$$= (3) \cdot \begin{bmatrix} 1 & \xi^2 & \xi & \xi^2 & \xi^2 & \xi^2 & \xi & \xi^2 & 1 \\ \xi & 1 & \xi^2 & \xi^2 & \xi^2 & \xi^2 & 1 & \xi & \xi^2 \\ \xi^2 & \xi & 1 & \xi^2 & \xi^2 & \xi^2 & \xi^2 & 1 & \xi \\ \xi & \xi^2 & 1 & 1 & \xi^2 & \xi & \xi^2 & \xi^2 & \xi^2 \\ 1 & \xi & \xi^2 & \xi & 1 & \xi^2 & \xi^2 & \xi^2 & \xi^2 \\ \xi^2 & 1 & \xi & \xi^2 & \xi & 1 & \xi^2 & \xi^2 & \xi^2 \\ \xi^2 & \xi^2 & \xi^2 & \xi & \xi^2 & 1 & 1 & \xi^2 & \xi \\ \xi^2 & \xi^2 & \xi^2 & 1 & \xi & \xi^2 & \xi & 1 & \xi^2 \\ \xi^2 & \xi^2 & \xi^2 & \xi^2 & 1 & \xi & \xi^2 & \xi & 1 \end{bmatrix} \qquad (26)$$

Introducing (26) in (24),

$$G = 3^{-2} W(2) C(2) S_f = W(2) \cdot 3^{-2} C(2) S_f = W(2) \cdot F. \qquad (27)$$

$$G = W(2) \cdot [\ 1\ \ 1\ \ 1\ \ 1\ \ \xi\ \ \xi^2\ \ 1\ \ \xi^2\ \ \xi\ ]^T. \qquad (28)$$

Recall that eq. (26) corresponds to $3^2 W(2)$. Hence

$$G = (1/3) \cdot \begin{bmatrix} 1 & \xi^2 & \xi & \xi^2 & \xi^2 & \xi^2 & \xi & \xi^2 & 1 \\ \xi & 1 & \xi^2 & \xi^2 & \xi^2 & \xi^2 & 1 & \xi & \xi^2 \\ \xi^2 & \xi & 1 & \xi^2 & \xi^2 & \xi^2 & \xi^2 & 1 & \xi \\ \xi & \xi^2 & 1 & 1 & \xi^2 & \xi & \xi^2 & \xi^2 & \xi^2 \\ 1 & \xi & \xi^2 & \xi & 1 & \xi^2 & \xi^2 & \xi^2 & \xi^2 \\ \xi^2 & 1 & \xi & \xi^2 & \xi & 1 & \xi^2 & \xi^2 & \xi^2 \\ \xi^2 & \xi^2 & \xi^2 & \xi & \xi^2 & 1 & 1 & \xi^2 & \xi \\ \xi^2 & \xi^2 & \xi^2 & 1 & \xi & \xi^2 & \xi & 1 & \xi^2 \\ \xi^2 & \xi^2 & \xi^2 & \xi^2 & 1 & \xi & \xi^2 & \xi & 1 \end{bmatrix} \cdot \begin{bmatrix} 1 \\ 1 \\ 1 \\ 1 \\ \xi \\ \xi^2 \\ 1 \\ \xi^2 \\ \xi \end{bmatrix} = \begin{bmatrix} \xi \\ 1 \\ \xi^2 \\ 1 \\ \xi \\ \xi^2 \\ \xi^2 \\ \xi^2 \\ \xi^2 \end{bmatrix} \qquad (29)$$

From
$$G = [\ \xi\ \ 1\ \ \xi^2\ \ 1\ \ \xi\ \ \xi^2\ \ \xi^2\ \ \xi^2\ \ \xi^2\ ]^T \qquad (30)$$
follows
$$g = [\ 1\ \ 0\ \ 2\ \ 0\ \ 1\ \ 2\ \ 2\ \ 2\ \ 2\ ]^T \qquad (31)$$
and $g = 2\ x_1 x_2 \oplus 2\ x_1 \oplus 2\ x_2 \oplus 1$. $\qquad (32)$

Since (32) comprises four spectral invariant operations applied to $x_1 x_2$ it follows that $g$ is also a bent function. Only one generalized permutation operation W(2) was used in the analysis based on the spectral domain. The calculation of W(2) was in this case complexer as in the former cases, since P(2) does not have a Kronecker product structure. Notice, the strong block structure of W(2), (as enhanced by the blue lines), similar to the one obtained in eq. (16).

Recall Z(1). In this case a generalized block-diagonal permutation matrix P'(2) may also be used.

$$P'(2) = blockdiag(\xi^2 Z(1),\ I(1),\ \xi Z^*(1)) \qquad (33)$$

It is however simple to show that the matrix P'(2) is identical with P(2), except that it has a different formal specification and this might alleviate the calculation of W(2). For example, P'(2) has an "additive Kronecker product decomposition", as follows:

$$P'(2) = (\xi^2 Z(1) \otimes diag(1\ 0\ 0)) + (I(1) \otimes diag(0\ 1\ 0)) + (\xi Z^*(1) \otimes diag(0\ 0\ 1)) \tag{34}$$

It is easy to see that now P'(2) is represented as the sum of three simple Kronecker products.

Following (9),

$$W'(2) = 3^{-2}C(2)P'(2)C^*(2) = 3^{-2}C(2)[(\xi^2 Z(1) \otimes diag(1\ 0\ 0)) + (I(1) \otimes diag(0\ 1\ 0)) + (\xi Z^*(1) \otimes diag(0\ 0\ 1))]C^*(2)$$

$$= 3^{-2}C(2)(\xi^2 Z(1) \otimes diag(1\ 0\ 0))C^*(2) + 3^{-2}C(2)(I(1) \otimes diag(0\ 1\ 0))C^*(2) + 3^{-2}C(2)(\xi Z^*(1) \otimes diag(0\ 0\ 1))C^*(2) \tag{35}$$

Since $C(2) = C(1) \otimes C(1)$, the first term of the former sum (35) may be expressed as:

$$3^{-1}(C(1) \otimes C(1))((\xi^2 Z(1) \otimes diag(1\ 0\ 0)))\ 3^{-1}(C^*(1) \otimes C^*(1))$$

which, taking advantage of the compatibility of the Kronecker product and the classical product of matrices [4], may be decomposed as

$$3^{-1}C(1)\ \xi^2 Z(1)\ C^*(1) \otimes 3^{-1}C(1)\ diag(1\ 0\ 0)\ C^*(1) \tag{36}$$

A similar decomposition may be done for the other terms of the sum (35).

$$3^{-2}C(2)(I(1) \otimes diag(0\ 1\ 0))C^*(2) = 3^{-1}C(1)\ I(1)\ C^*(1) \otimes 3^{-1}C(1)\ diag(0\ 1\ 0)\ C^*(1) \tag{37}$$

$$3^{-2}C(2)(\xi Z^*(1) \otimes diag(0\ 0\ 1))C^*(2) = 3^{-1}C(1)\ \xi Z^*(1)\ C^*(1) \otimes 3^{-1}C(1)\ diag(1\ 0\ 0)\ C^*(1) \tag{38}$$

Since $diag(1\ 0\ 0)$, $diag(0\ 1\ 0)$ and $diag(0\ 0\ 1)$ are independent of the other elementary permutations, then $3^{-1}C(1)\ diag(1\ 0\ 0)\ C^*(1)$, $3^{-1}C(1)\ diag(0\ 1\ 0)\ C^*(1)$, and $3^{-1}C(1)\ diag(0\ 0\ 1)\ C^*(1)$ should be calculated only once (and saved) to be used for the decomposition of other permutations with a blockdiagonal structure applied to ternary functions of two variables. Furthermore, as shown in equation (36), the matrix $3^{-1}C(1)\ diag(1\ 0\ 0)\ C^*(1)$ will be repeated at the positions corresponding to the non-zero entries of the permutation shown as first term of the Kronecker product, eventually scaled by a power of $\xi$. Similarly for $3^{-1}C(1)\ diag(0\ 1\ 0)\ C^*(1)$, and $3^{-1}C(1)\ diag(0\ 0\ 1)\ C^*(1)$.

$$3^{-1}C(1)\ diag(1\ 0\ 0)\ C^*(1) = 3^{-1}\begin{bmatrix} 1 & 1 & 1 \\ 1 & \xi & \xi^2 \\ 1 & \xi^2 & \xi \end{bmatrix} \cdot \begin{bmatrix} 1 & 0 & 0 \\ 0 & 0 & 0 \\ 0 & 0 & 0 \end{bmatrix} \cdot \begin{bmatrix} 1 & 1 & 1 \\ 1 & \xi^2 & \xi \\ 1 & \xi & \xi^2 \end{bmatrix} = 3^{-1}\begin{bmatrix} 1 & 1 & 1 \\ 1 & 1 & 1 \\ 1 & 1 & 1 \end{bmatrix} \tag{39}$$

$$3^{-1}C(1)\ diag(0\ 1\ 0)\ C^*(1) = 3^{-1}\begin{bmatrix} 1 & 1 & 1 \\ 1 & \xi & \xi^2 \\ 1 & \xi^2 & \xi \end{bmatrix} \cdot \begin{bmatrix} 0 & 0 & 0 \\ 0 & 1 & 0 \\ 0 & 0 & 0 \end{bmatrix} \cdot \begin{bmatrix} 1 & 1 & 1 \\ 1 & \xi^2 & \xi \\ 1 & \xi & \xi^2 \end{bmatrix} = 3^{-1}\begin{bmatrix} 1 & \xi^2 & \xi \\ \xi & 1 & \xi^2 \\ \xi^2 & \xi & 1 \end{bmatrix} \tag{40}$$

$$3^{-1}C(1)\ diag(0\ 0\ 1)\ C^*(1) = 3^{-1}\begin{bmatrix} 1 & 1 & 1 \\ 1 & \xi & \xi^2 \\ 1 & \xi^2 & \xi \end{bmatrix} \cdot \begin{bmatrix} 0 & 0 & 0 \\ 0 & 0 & 0 \\ 0 & 0 & 1 \end{bmatrix} \cdot \begin{bmatrix} 1 & 1 & 1 \\ 1 & \xi^2 & \xi \\ 1 & \xi & \xi^2 \end{bmatrix} = 3^{-1}\begin{bmatrix} 1 & \xi & \xi^2 \\ \xi^2 & 1 & \xi \\ \xi & \xi^2 & 1 \end{bmatrix} \tag{41}$$

Notice that (41) is the complex conjugate of (40)

The terms corresponding to the permutation components are:

$$3^{-1}C(1)\ \xi^2 Z(1)\ C^*(1) = 3^{-1}\xi^2\begin{bmatrix} 1 & 1 & 1 \\ 1 & \xi & \xi^2 \\ 1 & \xi^2 & \xi \end{bmatrix} \cdot \begin{bmatrix} 1 & 0 & 0 \\ 0 & \xi & 0 \\ 0 & 0 & \xi^2 \end{bmatrix} \cdot \begin{bmatrix} 1 & 1 & 1 \\ 1 & \xi^2 & \xi \\ 1 & \xi & \xi^2 \end{bmatrix} = \begin{bmatrix} 0 & \xi^2 & 0 \\ 0 & 0 & \xi^2 \\ \xi^2 & 0 & 0 \end{bmatrix} \tag{42}$$

$$3^{-1}C(1)\ I(1)\ C^*(1) = 3^{-1}C(1)C^*(1) = I(1) \tag{43}$$

$$3^{-1}C(1)\,\xi Z^*(1)\,C^*(1) = 3^{-1}\xi \begin{bmatrix} 1 & 1 & 1 \\ 1 & \xi & \xi^2 \\ 1 & \xi^2 & \xi \end{bmatrix} \cdot \begin{bmatrix} 1 & 0 & 0 \\ 0 & \xi^2 & 0 \\ 0 & 0 & \xi \end{bmatrix} \cdot \begin{bmatrix} 1 & 1 & 1 \\ 1 & \xi^2 & \xi \\ 1 & \xi & \xi^2 \end{bmatrix} = \begin{bmatrix} 0 & 0 & \xi \\ \xi & 0 & 0 \\ 0 & \xi & 0 \end{bmatrix} \quad (44)$$

Introducing the obtained matrices in equations (36), (37) and (38) leads to:

$$W(2) = 3^{-1} \left( \left( \begin{bmatrix} 0 & \xi^2 & 0 \\ 0 & 0 & \xi^2 \\ \xi^2 & 0 & 0 \end{bmatrix} \otimes \begin{bmatrix} 1 & 1 & 1 \\ 1 & 1 & 1 \\ 1 & 1 & 1 \end{bmatrix} \right) + \left( I(1) \otimes \begin{bmatrix} 1 & \xi^2 & \xi \\ \xi & 1 & \xi^2 \\ \xi^2 & \xi & 1 \end{bmatrix} \right) + \left( \begin{bmatrix} 0 & 0 & \xi \\ \xi & 0 & 0 \\ 0 & \xi & 0 \end{bmatrix} \otimes \begin{bmatrix} 1 & \xi & \xi^2 \\ \xi^2 & 1 & \xi \\ \xi & \xi^2 & 1 \end{bmatrix} \right) \right) \quad (45)$$

Notice that equation (45) present the same block structure as shown in the matrix of equation (29).

Let $P_{01}(1)$ denote the following permutation matrix:

$$P_{01}(1) = \begin{bmatrix} 0 & 1 & 0 \\ 1 & 0 & 0 \\ 0 & 0 & 1 \end{bmatrix}.$$

Consider now a generalized permutation matrix, which is different from P(2),

$$P''(2) = (\,\xi P_{12}(1) \otimes P_{01}(1)N(1)Z(1)\,) = \begin{bmatrix} \xi & 0 & 0 \\ 0 & 0 & \xi \\ 0 & \xi & 0 \end{bmatrix} \otimes \begin{bmatrix} 0 & \xi & 0 \\ 0 & 0 & \xi^2 \\ 1 & 0 & 0 \end{bmatrix} \quad (46)$$

It is fairly obvious that P''(2) is different from P(2).

Following (27)
$$W''(2) = C(2)P''(2)\langle C(2)\rangle^{-1} = 3^{-2}\,C(2)P''(2)C^*(2) = 3^{-2}\,C(2)(\,\xi P_{12}(1) \otimes P_{01}(1)N(1)Z(1)\,)C^*(2),$$

which taking again advantage of the compatibility of the Kronecker product and the classical product of matrices [4], may be decomposed as

$$W''(2) = 3^{-1}C(1)\,(\,\xi P_{12}(1)\,)C^*(1) \otimes 3^{-1}C(1)(\,P_{01}(1)N(1)Z(1)\,)C^*(1) =$$

$$= \begin{bmatrix} \xi & 0 & 0 \\ 0 & 0 & \xi \\ 0 & \xi & 0 \end{bmatrix} \otimes \begin{bmatrix} 0 & 1 & 0 \\ 0 & 0 & \xi^2 \\ \xi & 0 & 0 \end{bmatrix} = \xi P_{12}(1) \otimes Z^*(1)X^T(1) \quad (47)$$

The Kronecker product of two sparse matrices will clearly be different than the matrix in equation (29). A direct calculation of (47) gives the following matrix:

$$W''(2) = \begin{bmatrix} 0 & \xi & 0 & 0 & 0 & 0 & 0 & 0 & 0 \\ 0 & 0 & 1 & 0 & 0 & 0 & 0 & 0 & 0 \\ \xi^2 & 0 & 0 & 0 & 0 & 0 & 0 & 0 & 0 \\ 0 & 0 & 0 & 0 & 0 & 0 & 0 & \xi & 0 \\ 0 & 0 & 0 & 0 & 0 & 0 & 0 & 0 & 1 \\ 0 & 0 & 0 & 0 & 0 & 0 & \xi^2 & 0 & 0 \\ 0 & 0 & 0 & 0 & \xi & 0 & 0 & 0 & 0 \\ 0 & 0 & 0 & 0 & 0 & 1 & 0 & 0 & 0 \\ 0 & 0 & 0 & \xi^2 & 0 & 0 & 0 & 0 & 0 \end{bmatrix}, \quad (48)$$

from where

$$W''(2) \cdot [1\ 1\ 1\ 1\ \xi\ \xi^2\ 1\ \xi^2\ \xi]^T = [\xi\ 1\ \xi^2\ 1\ \xi\ \xi^2\ \xi^2\ \xi^2\ \xi^2]^T = G \quad (49)$$

Two quite different expressions P(2) and P"(2) for a generalized permutation matrix or, in other words, two different spectral invariant operations in the spectral domain were found to solve the same problem.

Notice that since P"(2) comprises the Kronecker product of two (generalized) elementary permutations, then W"(2) is also a generalized permutation matrix.

**Case 4: Booby trap**

Naïve intuitive generalizations are dangerous. Let $f_1$ and $f_2$ be 2-place ternary bent functions and let $S_{f1}$ and $S_{f2}$ be the Vilenkin-Chrestenson spectra of their sign representations.

Let

$$P(2) = (1/3)(\,diag(S_{f1})\,) \quad (50)$$

Is there a ternary bent function $g$ such that $S_g = P(2) \cdot S_{f_2}$. ?

The original request $S_g = P(2) \cdot S_{f_2}$ looks very much like Case 3 and Example 1, except that the matrix P(2) is a generalized diagonal permutation based on the spectrum of a bent function $f_1$, and another bent spectrum $S_{f_2}$ is given as reference.

**Example 4:**

Let $f_1 = [0\ 0\ 0\ 0\ 1\ 2\ 0\ 2\ 1]^T$ and $f_2 = [0\ 2\ 1\ 2\ 0\ 1\ 1\ 1\ 1]^T$. Both functions are ternary bent functions. A bent function $g$ is looked for, such that $S_g = (1/3)(diag(S_{f_1})) \cdot S_{f_2}$. Table 3 shows the preliminary calculation steps, from where it becomes apparent that $S_g$ is flat and all its components have magnitude 3, which corresponds to $3^{n/2}$, since $n = 2$.

If $S_g = P(2)\,(S_{f_2})$, then the inverse of $S_g$ should return $g$. It gives however the vector [0 0 0 0 0 0 0 0 $3\xi$ ]$^T$, which is not even the value vector of the sign of *any* ternary function. (No power of $\xi$ with a ternary exponent equals 0). This shows that the generalized permutation P(2) = $\frac{1}{3}(diag(S_{f_1}))$ is one of the many permutations that preserve the flatness of the spectrum of the bent reference function $f_2$, but does not produce a new bent function.

TABLE 3: PRELIMINARY CALCULATION STEPS

| $f_1$ | $f_2$ | $S_{f_1}$ | $S_{f_2}$ | diagonal of $P(2)$ | $S_g = P(2)\,(S_{f_2})$ |
|---|---|---|---|---|---|
| 0 | 0 | 3 | $3\xi$ | 1 | $3\xi$ |
| 0 | 2 | 3 | $3\xi^2$ | 1 | $3\xi^2$ |
| 0 | 1 | 3 | 3 | 1 | 3 |
| 0 | 2 | 3 | $3\xi^2$ | 1 | $3\xi^2$ |
| 1 | 0 | $3\xi^2$ | $3\xi$ | $\xi^2$ | 3 |
| 2 | 1 | $3\xi$ | 3 | $\xi$ | $3\xi$ |
| 0 | 1 | 3 | 3 | 1 | 3 |
| 2 | 1 | $3\xi$ | 3 | $\xi$ | $3\xi$ |
| 1 | 1 | $3\xi^2$ | 3 | $\xi^2$ | $3\xi^2$ |

## 4 GENERATING THE SET OF 2-PLACE TERNARY BENT FUNCTIONS IN THE SPECTRAL DOMAIN

It is known that using the classical 5 spectral invariant operations it is possible to partition the set of 2-place ternary bent functions into 9 equivalent classes of 54 functions each [17]. The first 18 functions of each class may be considered as "primitive" and the remaining 36 correspond to functions, which values equal those of the primitive functions shifted by 1 and by 2 modulo 3, respectively.

Lemma 1:

Let $r$ be a 2-place ternary bent function of a class and let $S_r$ be the Vilenkin-Chrestenson spectrum of its sign representation. This spectrum is flat. Moreover, let $g(x) = r(x) \oplus 1$, which is bent, since the addition of 1 is one of the classical spectral invariant operations. Then: $S_g = \xi \cdot S_r$.

Proof:
$$S_g = C^*(2)\xi^{g(x)} = C^* \xi^{r(x) \oplus 1} = \xi \cdot C^* \xi^{r(x)} = \xi \cdot S_r$$

From Lemma 1 follows that if $f$ is the representative of a class and P(2) is a permutation such that $S_r$ = P(2)·$S_f$. Then it becomes apparent that $S_g = \xi \cdot S_r = \xi \cdot$P(2)·$S_f$. It is clear that $\xi$·P(2) is a generalized permutation matrix relating the spectra of the signs of $g$ and $f$. It is simple to show that if $g(x) = r(x) \oplus 2$, then $S_g = \xi^2 \cdot S_r$.

**Theorem 1**:

Let $\Gamma$ = { I(1), $P_{01}$(1), $P_{12}$(1), N(1), X(1), $X^T$(1) } be the set of basic 3 × 3 permutations.
For each independent reference bent function "$f$" with sign representation $F$, being the "seed" of a class, as shown in Table 4, there are 18 different straight permutations P(2) = $\alpha \otimes \beta$, where $\alpha, \beta \in \Gamma$, $\alpha, \beta$ not necessarily different, leading to $S_g$ = P(2)$S_f$, from where 18 different $g$ functions with sign representation $G$ are obtained from the corresponding W(2)·$F$.

Remark:

I(1) $\otimes$ I(1) = I(2) does not constitute a real permutation in the sense of generating a "new" spectrum. Then there are 35 possible straight permutations P(2) = $\alpha \otimes \beta$, $\alpha, \beta \in \Gamma$. 18 of them are required to generate the permuted spectra and obtain the corresponding $g$ primitive functions. The remaining 17 P(2) permutations generate duplicates, but will not generate non-bent functions. For example if $f$ = [102 000 012]$^T$ and exponents($S_f$) = [001 211 121]$^T$ then both P(2) = X(1) $\otimes$ I(1) and $P_{01}$(1) $\otimes$ $P_{01}$(1) generate exponents(P(2)$S_f$) = [121 001 211]$^T$; both P(2) = $P_{01}$(1) $\otimes$ I(1) and X(1) $\otimes$ $P_{01}$(1) generate exponents(P(2)$S_f$) = [211 001 121]$^T$; both P(2) = X(1) $\otimes$ $X^T$(1) and $P_{01}$(1) $\otimes$ $P_{12}$(1) generate exponents(P(2)$S_f$) = [211 010 112]$^T$; both P(2) = N(1) $\otimes$ N(1) and $X^T$(1) $\otimes$ X(1) generate exponents(P(2)$S_f$) = [121 112 100]$^T$.

Proof: The proof of Theorem 1 is constructive. In the appendix, for each of the 9 independent classes, characterized by the reference functions of Table 4, the list of the 18 permuted spectra, the 18 P(2) permutations, the corresponding W(2) generalized permutations and the $g$ functions are given.

Corollary 1.1:

Recall from Lemma 1, that $S_{g \oplus 1} = \xi \cdot S_g$. Then $S_{g \oplus 1} = (\xi \cdot P(2)) \cdot S_f$ and $\xi^{g \oplus 1} = (\xi \cdot W(2)) \cdot F = \xi \cdot G$.

Corollary 1.1 indicates that rotating the P(2) permutation matrices of Theorem 1 by $\xi$ or $\xi^2$ generates functions that are obtained by adding 1 or 2, respectively, to the first 18 functions of each class thus completing the 54 functions of the corresponding full class.

Corollary 1.2:

All 486 2-place ternary bent functions are obtained starting with the Vilenkin-Chrestenson spectrum of 9 reference functions (Table 4); permutation matrices $P(2) = \alpha \otimes \beta$, $\alpha, \beta \in \Gamma$, $\alpha, \beta$ not necessarily different, and rotations as indicated in Corollary 1.1.

TABLE 4: REFERENCE BENT FUNCTIONS TO GENERATE ALL 2-PLACE TERNARY BENT FUNCTIONS

| Class | Reference function | exponents of $S_f$ |
|---|---|---|
| 1 | $f = [000\ 012\ 021]^T = x_1 x_2 \mod 3$ | $[000\ 021\ 012]^T$ |
| 2 | $f = [001\ 010\ 022]^T = x_1 x_2 \oplus x_2 \oplus 2(x_2)^2 \mod 3$ | $[000\ 021\ 120]^T$ |
| 3 | $f = [210\ 000\ 012]^T = x_1 x_2 \oplus (x_1)^2 \oplus 2x_2 \oplus 2 \mod 3$ | $[002\ 212\ 122]^T$ |
| 4 | $f = [100\ 010\ 220]^T = 2x_1 x_2 \oplus 2x_1 \oplus 2(x_2)^2 \oplus 1 \mod 3$ | $[012\ 021\ 111]^T$ |
| 5 | $f = [200\ 110\ 020]^T = 2x_1 x_2 \oplus 2x_1 \oplus (x_2)^2 \oplus 2 \mod 3$ | $[012\ 021\ 222]^T$ |
| 6 | $f = [102\ 000\ 012]^T = x_1 x_2 \oplus 2x_2 \oplus 2(x_1)^2 \oplus 1 \mod 3$ | $[001\ 211\ 121]^T$ |
| 7 | $f = [000\ 201\ 021]^T = x_1 x_2 \oplus x_1 \oplus (x_1)^2 \mod 3$ | $[020\ 011\ 002]^T$ |
| 8 | $f = [000\ 021\ 120]^T = 2x_1 x_2 \oplus x_1 \oplus 2(x_1)^2 \mod 3$ | $[010\ 022\ 001]^T$ |
| 9 | $f = [020\ 011\ 002]^T = 2x_1 x_2 \oplus x_2 \oplus (x_2)^2 \mod 3$ | $[000\ 012\ 210]^T$ |

Recall that as shown in (47) if a permutation P(2) is obtained as the Kronecker product of two 3×3 elementary permutations, then W(2) may also be decomposed as the Kronecker product of two 3×3 component permutations, which are generalized permutations, as shown below.

The basic cases are:

$$3^{-1}C(1)I(1)C^*(1) = 3^{-1}C(1)C^*(1) = I(1) \tag{51}$$

$$3^{-1}C(1)N(1)C^*(1) = 3^{-1}\begin{bmatrix} 1 & 1 & 1 \\ 1 & \xi & \xi^2 \\ 1 & \xi^2 & \xi \end{bmatrix} \cdot \begin{bmatrix} 0 & 0 & 1 \\ 0 & 1 & 0 \\ 1 & 0 & 0 \end{bmatrix} \cdot \begin{bmatrix} 1 & 1 & 1 \\ 1 & \xi^2 & \xi \\ 1 & \xi & \xi^2 \end{bmatrix} =$$

$$= 3^{-1}\begin{bmatrix} 1 & 1 & 1 \\ \xi^2 & \xi & 1 \\ \xi & \xi^2 & 1 \end{bmatrix} \cdot \begin{bmatrix} 1 & 1 & 1 \\ 1 & \xi^2 & \xi \\ 1 & \xi & \xi^2 \end{bmatrix} = \begin{bmatrix} 1 & 0 & 0 \\ 0 & 0 & \xi^2 \\ 0 & \xi & 0 \end{bmatrix} = Z^*(1) \cdot P_{12}(1) \tag{52}$$

$$3^{-1}C(1)P_{12}(1)C^*(1) = 3^{-1}\begin{bmatrix} 1 & 1 & 1 \\ 1 & \xi & \xi^2 \\ 1 & \xi^2 & \xi \end{bmatrix} \cdot \begin{bmatrix} 1 & 0 & 0 \\ 0 & 0 & 1 \\ 0 & 1 & 0 \end{bmatrix} \cdot \begin{bmatrix} 1 & 1 & 1 \\ 1 & \xi^2 & \xi \\ 1 & \xi & \xi^2 \end{bmatrix} =$$

$$= 3^{-1}\begin{bmatrix} 1 & 1 & 1 \\ 1 & \xi^2 & \xi \\ 1 & \xi & \xi^2 \end{bmatrix} \cdot \begin{bmatrix} 1 & 1 & 1 \\ 1 & \xi^2 & \xi \\ 1 & \xi & \xi^2 \end{bmatrix} = \begin{bmatrix} 1 & 0 & 0 \\ 0 & 0 & 1 \\ 0 & 1 & 0 \end{bmatrix} = P_{12}(1) \tag{53}$$

(Notice that $P_{12}(1)$ is trivially a generalized permutation, since $1 = \xi^0$.)

$$3^{-1}C(1)P_{01}(1)C^*(1) = 3^{-1}\begin{bmatrix} 1 & 1 & 1 \\ 1 & \xi & \xi^2 \\ 1 & \xi^2 & \xi \end{bmatrix} \cdot \begin{bmatrix} 0 & 1 & 0 \\ 1 & 0 & 0 \\ 0 & 0 & 1 \end{bmatrix} \cdot \begin{bmatrix} 1 & 1 & 1 \\ 1 & \xi^2 & \xi \\ 1 & \xi & \xi^2 \end{bmatrix} =$$

$$= 3^{-1}\begin{bmatrix} 1 & 1 & 1 \\ \xi & 1 & \xi^2 \\ \xi^2 & 1 & \xi \end{bmatrix} \cdot \begin{bmatrix} 1 & 1 & 1 \\ 1 & \xi^2 & \xi \\ 1 & \xi & \xi^2 \end{bmatrix} = \begin{bmatrix} 1 & 0 & 0 \\ 0 & 0 & \xi \\ 0 & \xi^2 & 0 \end{bmatrix} = Z(1) \cdot P_{12}(1) \tag{54}$$

Recall the Pauli matrix X(1) introduced in Example 1.

Then

$$3^{-1}C(1)\,X(1)\,C^*(1) = 3^{-1}\begin{bmatrix}1 & 1 & 1\\ 1 & \xi & \xi^2\\ 1 & \xi^2 & \xi\end{bmatrix}\cdot\begin{bmatrix}0 & 0 & 1\\ 1 & 0 & 0\\ 0 & 1 & 0\end{bmatrix}\cdot\begin{bmatrix}1 & 1 & 1\\ 1 & \xi^2 & \xi\\ 1 & \xi & \xi^2\end{bmatrix} =$$

$$= 3^{-1}\begin{bmatrix}1 & 1 & 1\\ \xi & \xi^2 & 1\\ \xi^2 & \xi & 1\end{bmatrix}\cdot\begin{bmatrix}1 & 1 & 1\\ 1 & \xi^2 & \xi\\ 1 & \xi & \xi^2\end{bmatrix} = \begin{bmatrix}1 & 0 & 0\\ 0 & \xi & 0\\ 0 & 0 & \xi^2\end{bmatrix} = Z(1) \quad (55)$$

and

$$3^{-1}C(1)\,X(1)^T\,C^*(1) = 3^{-1}\begin{bmatrix}1 & 1 & 1\\ 1 & \xi & \xi^2\\ 1 & \xi^2 & \xi\end{bmatrix}\cdot\begin{bmatrix}0 & 1 & 0\\ 0 & 0 & 1\\ 1 & 0 & 0\end{bmatrix}\cdot\begin{bmatrix}1 & 1 & 1\\ 1 & \xi^2 & \xi\\ 1 & \xi & \xi^2\end{bmatrix} =$$

$$3^{-1}\begin{bmatrix}1 & 1 & 1\\ \xi^2 & 1 & \xi\\ \xi & 1 & \xi^2\end{bmatrix}\cdot\begin{bmatrix}1 & 1 & 1\\ 1 & \xi^2 & \xi\\ 1 & \xi & \xi^2\end{bmatrix} = \begin{bmatrix}1 & 0 & 0\\ 0 & \xi^2 & 0\\ 0 & 0 & \xi\end{bmatrix} = Z^*(1) \quad (56)$$

It should be recalled, that for a given $p$, the tensor sum of two bent functions returns a new bent function on all the considered variables [11]. Hence the tensor sum of two 2-place ternary bent functions returns a 4-place ternary bent function.

Lemma 2:

Let $f_1$ and $f_2$ be 2-place ternary bent functions, with sign representations $F_1$ and $F_2$ and Vilenkin-Chrestenson spectra $S_{f_1}$ and $S_{f_2}$ respectively. From Theorem 1 it is known that if P'(2) = $\alpha_1 \otimes \beta_1$ and P''(2) = $\alpha_2 \otimes \beta_2$, with $\alpha_1$, $\beta_1$, $\alpha_2$, $\beta_2 \in \Gamma$, then there exist 2-place ternary bent functions $g_1$ and $g_2$ such that $S_{g_1}$ = P'(2) $S_{f_1}$ and $S_{g_2}$ = P''(2) $S_{f_2}$. Let $g_3 = g_1 \boxplus g_2$, where $\boxplus$ denotes the tensor sum. It is known [11], that $S_{g_3} = S_{g_1} \otimes S_{g_2}$. Then:

$$S_{g_3} = \left(P'(2)\,S_{f_1}\right) \otimes \left(P''(2)\,S_{f_2}\right) = (P'(2) \otimes P''(2))\cdot\left(S_{f_1} \otimes S_{f_2}\right) \quad (57)$$

Let $f_3 = f_1 \boxplus f_2$. Then $S_{f_3} = S_{f_1} \otimes S_{f_2}$. Moreover, let P(4) = P'(2) $\otimes$ P''(2). (58)

Introducing (58) in (57) follows that $S_{g_3}$ = P(4) $S_{f_3}$. (59)

A straight forward extension by $k$ repetitions of Lemma 2 shows that if $k \in \mathbb{N}$, $n = 2k$ is even, and let $f$ be an $n$-place ternary bent function with Vilenkin-Chrestenson spectrum $S_f$. Then if P($n$) is a permutation obtained as the Kronecker product of $n$ elementary permutations from $\Gamma$, $S_g$ = P($n$) $S_f$ is the spectrum of an $n$-place ternary bent function $g$.

Theorem 1 and the extension of Lemma 2 prove the following:

**Theorem 2**:

Let $f_1$ and $f_2$ be $n$-place and $m$-place ternary bent functions, with $m$ and $n$ even, sign representations $F_1$ and $F_2$ and Vilenkin-Chrestenson spectra $S_{f_1}$ and $S_{f_2}$ respectively. Let $f_3 = f_1 \boxplus f_2$, from where $f_3$ is an ($n+m$)-place ternary bent function. Then $S_{f_3} = S_{f_1} \otimes S_{f_2}$. Let P($n$) and P($m$) be permutations obtained with the Kronecker product of $n$ and $m$ elementary permutations of $\Gamma$. Let $S_{g_1}$ = P($n$) $S_{f_1}$ and $S_{g_2}$ = P($m$) $S_{f_2}$, which from the extension of Theorem 1 will be the spectra of an $n$-place and an $m$-place ternary bent functions respectively. If $g_3$ is defined to be equal to $g_1 \boxplus g_2$, then $g_3$ will be an ($n+m$)-place ternary bent function, and

$$S_{g_3} = S_{g_1} \otimes S_{g_2} = (P(n)\, S_{f_1}) \otimes (P(m)\, S_{f_2}) = (P(n) \otimes P(m))\, (S_{f_1} \otimes S_{f_2}) = P(n+m)\, (S_{f_1} \otimes S_{f_2}) = P(n+m)\, S_{f_3}. \tag{60}$$

The extension of Theorem 1 and Theorem 2 prove that a spectral permutation based on Γ, applied to the spectrum of a ternary bent function of an even number $n$ of variables, obtained with a tensor sum of two component bent functions on $(n/2)$ variables, produces the spectrum of another ternary bent function on the same number $n$ of variables.

The fact that Theorems 1 and 2 use an even number of variables opens the question whether in the case of ternary bent functions generated with the Maiorana method [7], [12], which requires an even number of variables and produces strict bent functions, an analoge to the extension of Lemma 2 exists. It should be recalled, that Maiorana bent functions and bent functions obtained as the tensor sum of two bent functions, are disjunct [12].

This leads to:

**Theorem 3**:

If $n$ is even and $f$ is an $n$-place ternary bent Maiorana function with sign representation $F$, and Vilenkin-Chrestenson spectrum $S_f$, then there is an $n$-place ternary bent function $g$ such that $S_g = P(n)S_f$, where $P(n)$ is a permutation comprising the Kronecker product of $n$ elementary permutations of Γ.

The following notation and facts will be used in the proof:

Recall that $h: \{0, 1, 2\} \leftrightarrow \{1, \xi, \xi^2\}$. For all $v_1, v_2 \in \mathbb{Z}_3$, $h(v_1) \cdot h(v_2) = h(v_1 \oplus v_2)$.

Let $M = [m_{i,j}]$, where $m_{i,j} = \xi^{i \cdot j\ mod\ 3}$, $i, j \in \mathbb{Z}_3$. Then $C(1) = h(M)$.

$M^{[r]} = M \boxplus M \boxplus \ldots \boxplus M$, $r$ times. Then $C(r) = h(M^{[r]})$ [12]

$Q(n)$ denotes a $p^n \times p^n$ straight permutation matrix. $1_{3^k}$ denotes a column vector of length $3^k$, $k \leq n$, with all entries equal to 1.

Let $A(n)$ be an arbitrary square matrix. Then $vec(A(n))$ denotes a vectorizing operation producing a vector of length $p^{2n}$, whose (ordered) cofactors are the columns of $A(n)$.

Let $v(n)$ denote the value vector of a random $n$-place ternary function.

Proof of the Theorem:

Let $n = 2m$, $m \in \mathbb{N}$. If the value vector of an $n$-place ternary function $f$ has the structure shown below, then $f$ is a Maiorana bent function [12].

$$f = vec\langle M^{[m]} \cdot Q(m) \oplus (1_{3^m} \otimes v^T(m)) \rangle \tag{61}$$

$$F = h(f) = h\,(vec\langle M^{[m]} \cdot Q(m) \oplus (1_{3^m} \otimes v^T(m)) \rangle) \tag{62}$$

Since $h$ projects ternary values onto the unit circle and $vec$ reorders the columns of a matrix into a vector without changing the values of the entries, $h$ applied to a vectorized matrix gives the same result as vectorizing the matrix obtained after applying $h$.

Therefore, following [12]:

$$F = vec\,(h\,\langle M^{[m]} \cdot Q(m) \oplus (1_{3^m} \otimes v^T(m)) \rangle) \tag{63}$$

$$F = vec\,(h\,\langle M^{[m]} \cdot Q(m) \rangle\, \#\, h(1_{3^m} \otimes v^T(m))) \tag{64}$$

where # denotes the Hadamard product of matrices [4].

$$F = vec\,(\,h(M^{[m]}) \cdot Q(m)\, \#\, (1_{3^m} \otimes h(v^T(m)))) \tag{65}$$

$$F = vec\left( C(m) \cdot Q(m) \cdot \begin{bmatrix} h(v(0)) & \cdots & 0 \\ \vdots & \ddots & \vdots \\ 0 & \cdots & h(v(3^m - 1)) \end{bmatrix} \right) \tag{66}$$

Notice that the entries of the diagonal matrix are powers of $\xi$, with exponents from $\mathbb{Z}_3$. The product of the diagonal matrix and Q($m$) will transform Q($m$) from a straight permutation into a generalized permutation Q'($m$). Since all entries of C($m$) are also powers of $\xi$, with exponents from $\mathbb{Z}_3$ (recall that $1 = \xi^0$), then the product C($m$)·Q'($m$) will return a matrix all whose entries are powers of $\xi$. Hence, $F$ is a vector all whose components are powers of $\xi$. The length of $F$ is $3^n$ since the matrix which is vectorized is of dimension $3^m \times 3^m$.

Now let the question be analyzed, whether there exists a ternary bent function $g$ of $n$ variables, such that the spectrum $S_g$ of its sign representation $G$, equals P($n$)$S_f$, where $S_f$ denotes the spectrum of $F$ and P($n$) corresponds to a straight permutation obtained as the Kronecker product of $n$ elementary permutations from $\Gamma$.

Since $f$ is a bent function, its spectrum is flat. All spectral coefficients have absolute value equal to $3^{n/2}$ and any permutation of a flat spectrum preserves the flatness, however not the spectral character.

Recall Eqs. (9) to (11):

$$W(n) = p^{-n}C(n)P(n)C^*(n) \quad \text{and} \quad G = W(n)F$$

Since it is clear that $S_g$ is flat, it is only necessary to prove that $g$ is a ternary function, with sign representation $G$, to have a bent function.

Recall that since P($n$) has a Kronecker product structure, then W($n$) may be decomposed into the Kronecker product of elementary generalized permutations (recall eqs. (47), (51)-(56)), which shows that W($n$) is a generalized permutation matrix.

Applying W($n$) to eq. (66) leads to:

$$G = W(n)F = W(n) \cdot vec\left( C(m) \cdot Q(m) \cdot \begin{bmatrix} h(v(0)) & \cdots & 0 \\ \vdots & \ddots & \vdots \\ 0 & \cdots & h(v(3^m - 1)) \end{bmatrix} \right) \quad (67)$$

As mentioned above, $F$ is a vector such that every component is a power of $\xi$. Since W($n$) is a generalized permutation, then $G$ is also a vector with all components being powers of $\xi$. This means that G is the sign representation of a *ternary* function g. Since the spectrum $S_g$ is flat, then $g$ is bent.

∎∎∎

It has been shown [12] that if $n = 2$, there are 156 ternary Maiorana bent functions. If $n > 2$, the number of Maiorana bent functions is not known, but it is known, that all of them are strict bent [12].

Let $n = 2m$, $m \in \mathbb{N}$. Given an $n$-place Maiorana bent function $f$, with (61) one has:

$$f = vec\langle M^{[m]} \cdot Q(m) \oplus (1_{3^m} \otimes v^T(m)) \rangle$$

With (3) and (6) $\quad S_f = C^*(n) \cdot h(f)$

and with Lemma 4.2.1 of [4],

$S_f = vec \langle C^*(m) \cdot vec^{-1}(h(f)) \cdot C^*(m) \rangle$

$\quad = vec \langle C^*(m) \cdot h(M^{[m]} \cdot Q(m) \oplus (1_{3^m} \otimes [\, v(0), v(1), \ldots, v(3^m-1)]^T) \cdot C^*(m) \rangle$

$\quad = vec \langle C^*(m) \cdot ((h(M^{[m]})) \cdot Q(m) \,\#\, (1_{3^m}) \otimes h([\, v(0), v(1), \ldots, v(3^m-1)]^T)) \cdot C^*(m) \rangle$

$\quad = vec \langle C^*(m) \cdot (C(m) \cdot Q(m) \cdot diag(\, h(v(0)), h(v(1), \ldots h(v(3^m-1)) \cdot C^*(m) \rangle$

$\quad = vec \langle p^m I(m) \cdot Q(m) \cdot diag(\, h(v(0)), h(v(1)), \ldots, h(v(3^m-1)) \cdot C^*(m) \rangle$

$\quad = p^m \cdot vec \langle Q(m) \cdot diag(\, h(v(0)), h(v(1), h(v(3^m-1)) \cdot C^*(m) \rangle$

Notice that the diagonal matrix will change Q($m$) into a generalized permutation, and the following product with C*($m$) will result in a matrix, all entries of which are powers of $\xi$.

Therefore, after vectorizing,

$$S_f = p^m \cdot [\xi^{t(0)}, \xi^{t(1)}, \ldots, \xi^{t(3^m-1)}]^T = p^m \cdot \xi^{t(0)}[1, \xi^{t(1)-t(0)}, \xi^{t(2)-t(0)}, \ldots, \xi^{t(3^m-1)-t(0)}]^T =$$
$$= [s(0), s(1), \ldots, s((3^m-1))]^T$$

It may be seen that $s(0) = p^m \cdot \xi^{t(0)}$ and all other spectral coefficients are rotations of $s(0)$ by $p^m$ times some power of $\xi$. This is a characterizing property of strict bent functions. Moreover, all spectral coefficients have absolute value equal to $p^m$, which corresponds to $p^{n/2}$, as required for bent functions. The function $t$ is known as the dual of $f$ and it is also strict bent [12].

The former analysis showed that if $f$ is an $n$-place Maiorana ternary bent function, with sign representation $F$, Vilkenkin-Chrestenson spectrum $S_f$, and P($n$) is a straight permutation obtained as the Kronecker product of $n$ elementary permutations of $\Gamma$, then P($n$)$S_f$ produces the spectrum of another $n$-place ternary bent function. Still to be proven is the question whether for any $n$ even, $> 2$, the obtained new $n$-place ternary bent function is also a Maiorana function or not. A negative answer exists for $n = 2$: A Maiorana ternary bent function is strict bent, and in the case of $n = 2$ it is known that all 468 ternary bent functions are strict bent [11], however only 156 are Maiorana bent functions [12].

A different structure of permutations will be considered in what follows.

Recall that eq. (33) introduced a permutation P(2) with a blockdiagonal structure, which solved the problem of Case 3. Moreover eqs. (34) to (45) showed a way of decomposing P(2) to simplify the calculation of W(2). It seems reasonable to ask whether it is possible to generate all 2-place ternary bent functions based on a set of reference functions and permutations with a blockdiagonal structure comprising elementary permutations from $\Gamma$.

The first preliminary question is related to the number of P(2) permutations with blockdiagonal structure, based on elementary permutations of $\Gamma$. If all 3 elementary permutations are chosen to be different, there are 6·5·4 = 120 choices, which may be reordered in 3! ways leading to 720 P(2) permutations. If only 2 elementary permutations are considered –(one will be included twice)- there are 6·5 = 30 choices, which may be reordered in 3 ways leading to additional 90 P(2) permutations. Finally if all 3 elementary permutations are equal, there are 5 choices (since it does not make sense to chose the identity). This amounts to a total of 815 possible P(2) permutations, from which only 18 would be needed if 9 classes are considered.

For example, let $f = [000\ 012\ 021]^T = x_1 x_2 \bmod 3$, which is bent, with exponents($S_f$) = $[000\ 021\ 012]^T$. It is simple to see that P(2) = blockdiag(I(1), I(1), X(1)) applied to $[000\ 021\ 012]^T$ gives $[000\ 021\ 201]^T$, which is bent and since all 2-place ternary bent functions are strict bent [11], then the corresponding $g$ function is also bent. It is simple to see that for the given $f$, P(2) = blockdiag(*any*, I(1), X(1)) will generate duplicates of the former solution. However if P(2) is chosen to be blockdiag(I(1), I(1), P$_{12}$) and is applied to the same reference spectrum, $[000\ 021\ 021]^T$ will be obtained, which is *not* bent, since it is known that if a 2-place ternary function is bent, and has a constant cofactor in its value vector, the other two cofactors are not equal. This example shows that if the spectrum of a reference bent function is permuted with a matrix with a block-diagonal structure, there is no guarantee that the obtained spectrum will be that of another bent function, in spite of being flat.

Recall that in a generalized permutation matrix, the non-zero entries may be scaled by -1. Then $-$ I($n$) is possibly one of the simplest generalized permutation matrices.

**Theorem 4**: Let $p$ be **odd** and let $f$ be a $p$-valued bent function with sign $F$ and spectrum $S_f$. Then if P(2) = $-$ I(2) there **does not exist** a $p$-valued bent function $g$ with sign $G$ such that $S_g$ = P(2)·$S_f$ = - $S_f$

Proof:

Should $S_g = -S_f$ hold, then $G = p^{-n} C(n) \cdot (-S_f) = -p^{-n} C(n) \cdot S_f = -F$.

Notice that $-F = -[\xi^{f(0)}, \xi^{f(1)}, \ldots, \xi^{f(p^n-1)}] = [-\xi^{f(0)}, -\xi^{f(1)}, \ldots, -\xi^{f(p^n-1)}]$

Let $a \in \mathbb{Z}_p^n$. Then

$$-\xi^{f(a)} = e^{i\pi} \cdot \xi^{f(a)} = e^{i\pi} \cdot e^{\frac{i2\pi f(a)}{p}} = e^{i\pi\left(\frac{2f(a)}{p}+1\right)} = e^{i2\pi\left(\frac{f(a)}{p}+\frac{1}{2}\right)} = e^{\frac{i2\pi}{p}(f(a)+\frac{p}{2})} = \xi^{\left(f(a)+\frac{p}{2}\right)}.$$

It is simple to see that if $p$ is odd, then $\left(f(a)+\frac{p}{2}\right)$ is not in $\mathbb{Z}_p$.

This means that $-F$ is not even the sign representation of a $p$-valued function.

Consider the case of $-\xi^1$ when $p = 3$ or $p = 5$:

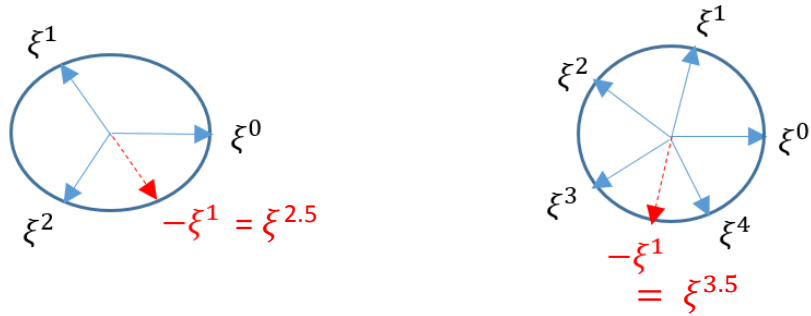

It is simple to see that $-\xi^1 = \xi^{1+p/2}$ is not the sign representation of a ternary or a 5-valued variable, whereas when e.g. $p = 4$ or $6$, the negative scaling of the powers of $\xi$ induces a permutation in the value set, shifting them by $p/2 \mod p$.

Consider the case of $-\xi^1$ when $p = 4$ or $p = 6$:

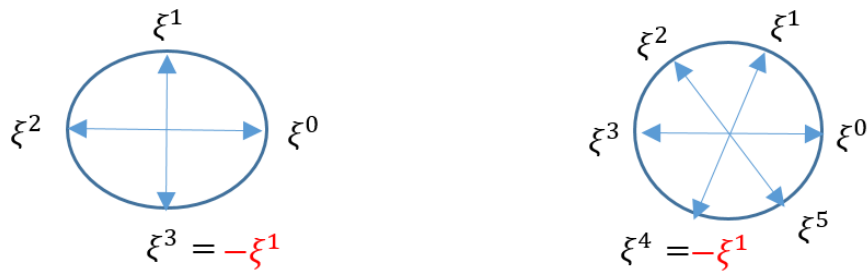

It becomes apparent that in these cases, a permutation of the value set takes place.

# 5 CLOSING REMARKS

Different aspects of the generation of ternary bent functions by applying permutations of their Vilenkin-Chrestenson spectra were studied. Since the value vector of an *n*-place ternary bent function as well as of its spectrum is of length $3^n$, there are $(3^n)!$ possible straight permutations. This number is much larger than the known number of ternary bent functions. For the simplest case of *n* = 2, it is known that there are 486 ternary bent functions, whereas $(3^2)!$ equals 362,880.

Therefore a focus of this research was the search for classes of *effective* permutations leading to new bent functions, the characterization of *weak* permutations leading to a small number of new bent functions and albeit, a large number of non-bent functions, and the characterization of classes of *bad* permutations leading only to non-bent functions.

It was shown that the class of permutations obtained as the Kronecker product of elementary permutations of the set $\Gamma$, is effective, producing new bent functions from known ones, however including duplicates, d.h. several different permutations may lead to a same bent function. With a set of 9 independent 2-place ternary bent functions and permutations of this class, all 2-place ternary bent functions could be generated. (See the Appendix).

It was also shown that the class of permutations with a diagonal block structure comprising elementary permutations of $\Gamma$ is weak, producing more non-bent functions than new ones.

Finally, a class was identified, which for any *p* odd, is bad. The inverse Vilenkin-Chrestenson spectrum of a permuted spectrum does not return a ternary function!

# Appendix

## 9 Classes of 2-place ternary bent functions based on spectral permutations

The classes are tabulated based on 5 columns. The first column enumerates the basic 18 functions of each class, where the reference function is assigned the number 1. The second column shows for each function the elements of the value vectors $g$ and $G$. Columns 3 and 4 present the P(2) and W(2) permutation matrices associated to each spectrum and function. The last column presents two views of the permuted spectra: the upper representation corresponds to the exponents of the spectral coefficients, whereas the lower one shows (for simplicity) the spectral coefficients without the scaling factor $3^{n/2}$ = 3. The notation $[\![S_g = \mathrm{P}(2)S_f]\!]$ at the head of the column is used to denote the spectra without the scaling factor.

Not shown are the functions obtained by adding 1 or 2 modulo 3 to the basic functions, which implies rotating the corresponding spectra by $\xi$ or $\xi^2$, respectively. In this way each class comprises 54 functions. The 9 classes cover then all 486 ternary bent functions of two variables.

# Classes of 2-place ternary bent functions

Class 1: Reference function $f$ = [0 0 0 0 1 2 0 2 1]$^T$ = $x_1 x_2 \; mod \; 3$;

$F$ = [ 1 1 1 1 $\xi$ $\xi^2$ 1 $\xi^2$ $\xi$ ]$^T$ ; $S_f$ = 3·[ 1 1 1 1 $\xi^2$ $\xi$ 1 $\xi$ $\xi^2$ ]$^T$ ; exponents of $S_f$ = [000 021 012]

| | $g$<br>$G = W(2) \cdot F$ | | | $[\![ S_g = P(2) S_f ]\!]$ |
|---|---|---|---|---|
| 2 | 000 021 012<br>1 1 1 1 $\xi$ $\xi^2$ 1 $\xi$ $\xi^2$ | P(2)<br>W(2) | $P_{12}(1) \otimes I(1)$<br>$P_{12}(1) \otimes I(1)$ | 000 012 021<br>1 1 1 1 $\xi^2$ $\xi$ 1 $\xi^2$ $\xi$ |
| 3 | 000 102 201<br>1 1 1 $\xi$ 1 $\xi^2$ $\xi^2$ 1 $\xi$ | P(2)<br>W(2) | $P_{01}(1) \otimes I(1)$<br>$Z(1)P_{12}(1) \otimes I(1)$ | 021 000 012<br>1 $\xi^2$ $\xi$ 1 1 1 1 $\xi$ $\xi^2$ |
| 4 | 000 120 210<br>1 1 1 $\xi^2$ 1 $\xi$ $\xi^2$ $\xi$ 1 | P(2)<br>W(2) | $X(1) \otimes I(1)$<br>$Z(1) \otimes I(1)$ | 012 000 021<br>1 $\xi$ $\xi^2$ 1 1 1 1 $\xi^2$ $\xi$ |
| 5 | 000 201 102<br>1 1 1 $\xi^2$ 1 $\xi$ $\xi$ 1 $\xi^2$ | P(2)<br>W(2) | $X^T(1) \otimes I(1)$<br>$Z^*(1) \otimes I(1)$ | 021 012 000<br>1 $\xi^2$ $\xi$ 1 $\xi$ $\xi^2$ 1 1 1 |
| 6 | 000 210 120<br>1 1 1 $\xi^2$ $\xi$ 1 $\xi$ $\xi^2$ 1 | P(2)<br>W(2) | $N(1) \otimes I(1)$<br>$Z^*(1)P_{12}(1) \otimes I(1)$ | 012 021 000<br>1 $\xi$ $\xi^2$ 1 $\xi^2$ $\xi$ 1 1 1 |
| 7 | 012 000 021<br>1 $\xi$ $\xi^2$ 1 1 1 1 $\xi^2$ $\xi$ | P(2)<br>W(2) | $I(1) \otimes P_{01}(1)$<br>$I(1) \otimes Z(1)P_{12}(1)$ | 000 201 102<br>1 1 1 $\xi^2$ 1 $\xi$ $\xi$ 1 $\xi^2$ |
| 8 | 012 021 000<br>1 $\xi$ $\xi^2$ 1 $\xi^2$ $\xi$ 1 1 1 | P(2)<br>W(2) | $I(1) \otimes X(1)$<br>$I(1) \otimes Z(1)$ | 000 102 201<br>1 1 1 $\xi$ 1 $\xi^2$ $\xi^2$ 1 $\xi$ |
| 9 | 012 102 222<br>1 $\xi$ $\xi^2$ $\xi$ 1 $\xi^2$ $\xi^2$ $\xi^2$ $\xi^2$ | P(2)<br>W(2) | $P_{01}(2)$<br>$Z(2)P_{12}(2)$ | 201 000 102<br>$\xi^2$ 1 $\xi$ 1 1 1 $\xi$ 1 $\xi^2$ |
| 10 | 012 111 210<br>1 $\xi$ $\xi^2$ $\xi$ $\xi$ $\xi$ $\xi^2$ $\xi$ 1 | P(2)<br>W(2) | $P_{12}(1) \otimes X(1)$<br>$Z(2) \cdot ( P_{12}(1) \otimes I(1))$ | 102 000 201<br>$\xi$ 1 $\xi^2$ 1 1 1 $\xi^2$ 1 $\xi$ |
| 11 | 012 210 111<br>1 $\xi$ $\xi^2$ $\xi^2$ $\xi$ 1 $\xi$ $\xi$ $\xi$ | P(2)<br>W(2) | $N(1) \otimes P_{01}(1)$<br>$(Z^*(1) \otimes Z(1))P_{12}(2)$ | 102 201 000<br>$\xi$ 1 $\xi^2$ $\xi^2$ 1 $\xi$ 1 1 1 |
| 12 | 012 222 102<br>1 $\xi$ $\xi^2$ $\xi^2$ $\xi^2$ $\xi^2$ $\xi$ 1 $\xi^2$ | P(2)<br>W(2) | $N(1) \otimes X(1)$<br>$Z^*(1)P_{12}(1) \otimes Z(1)$ | 201 102 000<br>$\xi^2$ 1 $\xi$ $\xi$ 1 $\xi^2$ 1 1 1 |
| 13 | 021 000 012<br>1 $\xi^2$ $\xi$ 1 1 1 1 $\xi$ $\xi^2$ | P(2)<br>W(2) | $I(1) \otimes X^T(1)$<br>$I(1) \otimes Z^*(1)$ | 000 210 120<br>1 1 1 $\xi^2$ $\xi$ 1 $\xi$ $\xi^2$ 1 |
| 14 | 021 012 000<br>1 $\xi^2$ $\xi$ 1 $\xi$ $\xi^2$ 1 1 1 | P(2)<br>W(2) | $I(1) \otimes N(1)$<br>$I(1) \otimes Z^*(1)P_{12}(1)$ | 000 120 210<br>1 1 1 $\xi$ $\xi^2$ 1 $\xi^2$ $\xi$ 1 |
| 15 | 021 111 201<br>1 $\xi^2$ $\xi$ $\xi$ $\xi$ $\xi$ $\xi^2$ 1 $\xi$ | P(2)<br>W(2) | $P_{01}(1) \otimes N(1)$<br>$(Z(1) \otimes Z^*(1)) \cdot P_{12}(2)$ | 120 000 210<br>$\xi$ $\xi^2$ 1 1 1 1 $\xi^2$ $\xi$ 1 |
| 16 | 021 120 222<br>1 $\xi^2$ $\xi$ $\xi$ $\xi^2$ 1 $\xi^2$ $\xi^2$ $\xi^2$ | P(2)<br>W(2) | $P_{01}(1) \otimes X^T(1)$<br>$Z(1)P_{12}(1) \otimes Z^*(1)$ | 210 000 120<br>$\xi^2$ $\xi$ 1 1 1 1 $\xi$ $\xi^2$ 1 |
| 17 | 021 201 111<br>1 $\xi^2$ $\xi$ $\xi^2$ 1 $\xi$ $\xi$ $\xi$ $\xi$ | P(2)<br>W(2) | $X^T(1) \otimes N(1)$<br>$Z^*(2) \cdot (I(1) \otimes P_{12}(1))$ | 120 210 000<br>$\xi$ $\xi^2$ 1 $\xi^2$ $\xi$ 1 1 1 1 |
| 18 | 021 222 120<br>1 $\xi^2$ $\xi$ $\xi^2$ $\xi^2$ $\xi^2$ $\xi$ $\xi^2$ 1 | P(2)<br>W(2) | $N(2)$<br>$Z^*(2)P_{12}(2)$ | 210 120 000<br>$\xi^2$ $\xi$ 1 $\xi$ $\xi^2$ 1 1 1 1 |

Class 2. Reference function $f = [\ 0\ 0\ 1\ 0\ 1\ 0\ 0\ 2\ 2\ ]^T = x_1 x_2 \oplus x_2 \oplus 2(x_2)^2 \ mod\ 3$;
$F = [\ 1\ 1\ \xi\ 1\ \xi\ 1\ 1\ \xi^2\ \xi^2\ ]^T$; $S_f = 3\cdot[\ 1\ 1\ 1\ 1\ \xi^2\ \xi\ \xi\ \xi^2\ 1]^T$; exponents of $S_f$ = [000 021 120]

| | $g$<br>$G = W(2)\cdot F$ | | | $[\![S_g = P(2)S_f]\!]$ |
|---|---|---|---|---|
| 2 | 001 022 010<br>1 1 $\xi$ 1 $\xi^2$ $\xi^2$ 1 $\xi$ 1 | P(2)<br>W(2) | $P_{12}(1) \otimes I(1)$<br>$P_{12}(1) \otimes I(1)$ | 000 120 021<br>1 1 1 $\xi^2$ 1 1 $\xi^2$ $\xi$ |
| 3 | 001 100 202<br>1 1 $\xi$ 1 1 $\xi^2$ 1 $\xi^2$ | P(2)<br>W(2) | $P_{01}(1) \otimes I(1)$<br>$Z(1)P_{12}(1) \otimes I(1)$ | 021 000 120<br>1 $\xi^2$ $\xi$ 1 1 1 $\xi$ $\xi^2$ 1 |
| 4 | 001 121 211<br>1 1 $\xi$ $\xi$ $\xi^2$ $\xi$ $\xi^2$ $\xi$ $\xi$ | P(2)<br>W(2) | $X(1) \otimes I(1)$<br>$Z(1) \otimes I(1)$ | 120 000 021<br>$\xi$ $\xi^2$ 1 1 1 1 1 $\xi^2$ $\xi$ |
| 5 | 001 202 100<br>1 1 $\xi^2$ 1 $\xi^2$ $\xi$ 1 1 | P(2)<br>W(2) | $X^T(1) \otimes I(1)$<br>$Z^*(1) \otimes I(1)$ | 021 120 000<br>1 $\xi^2$ $\xi$ $\xi$ $\xi^2$ 1 1 1 1 |
| 6 | 001 211 121<br>1 1 $\xi^2$ $\xi$ $\xi$ $\xi$ $\xi^2$ $\xi$ | P(2)<br>W(2) | $N(1) \otimes I(1)$<br>$Z^*(1)P_{12}(1) \otimes I(1)$ | 120 021 000<br>$\xi$ $\xi^2$ 1 1 $\xi^2$ $\xi$ 1 1 1 |
| 7 | 010 001 022<br>1 $\xi$ 1 1 1 $\xi$ 1 $\xi^2$ $\xi^2$ | P(2)<br>W(2) | $I(1) \otimes P_{12}(1)$<br>$I(1) \otimes P_{12}(1)$ | 000 012 102<br>1 1 1 $\xi$ 1 $\xi^2$ 1 $\xi$ $\xi^2$ |
| 8 | 010 022 001<br>1 $\xi$ 1 1 $\xi^2$ $\xi^2$ 1 1 $\xi$ | P(2)<br>W(2) | $I(1) \otimes X(1)$<br>$I(1) \otimes Z(1)$ | 000 102 012<br>1 1 1 $\xi$ $\xi^2$ 1 $\xi$ $\xi^2$ |
| 9 | 010 100 220<br>1 $\xi$ 1 $\xi$ 1 1 $\xi^2$ $\xi^2$ 1 | P(2)<br>W(2) | $P_{01}(1) \otimes P_{12}(1)$<br>$(Z(1) \otimes I(1))\cdot P_{12}(2)$ | 012 000 102<br>1 $\xi$ $\xi^2$ 1 1 1 $\xi$ 1 $\xi^2$ |
| 10 | 010 112 211<br>1 $\xi$ 1 $\xi$ $\xi$ $\xi^2$ $\xi^2$ $\xi$ $\xi$ | P(2)<br>W(2) | $P_{01}(1) \otimes X(1)$<br>$Z(1) \cdot P_{12}(1) \otimes Z(1)$ | 102 000 012<br>$\xi$ 1 $\xi^2$ 1 1 1 1 $\xi$ $\xi^2$ |
| 11 | 010 211 112<br>1 $\xi$ 1 $\xi^2$ $\xi$ $\xi$ $\xi$ $\xi$ $\xi^2$ | P(2)<br>W(2) | $N(1) \otimes P_{12}(1)$<br>$(Z^*(1) \otimes I(1))P_{12}(2)$ | 102 012 000<br>$\xi$ 1 $\xi^2$ 1 $\xi$ $\xi^2$ 1 1 1 |
| 12 | 010 220 100<br>1 $\xi$ 1 $\xi^2$ $\xi^2$ 1 $\xi$ 1 1 | P(2)<br>W(2) | $N(1) \otimes X(1)$<br>$Z(1)^*P_{12}(1) \otimes Z(1)$ | 012 102 000<br>1 $\xi$ $\xi^2$ $\xi$ 1 $\xi^2$ 1 1 1 |
| 13 | 022 001 010<br>1 $\xi^2$ $\xi^2$ 1 1 $\xi$ 1 $\xi$ 1 | P(2)<br>W(2) | $I(1) \otimes X^T(1)$<br>$I(1) \otimes Z^*(1)$ | 000 210 201<br>1 1 1 $\xi^2$ $\xi$ 1 $\xi^2$ 1 $\xi$ |
| 14 | 022 010 001<br>1 $\xi^2$ $\xi^2$ 1 $\xi$ 1 1 1 $\xi$ | P(2)<br>W(2) | $I(1) \otimes P_{01}(1)$<br>$I(1) \otimes Z(1)P_{12}(1)$ | 000 201 210<br>1 1 1 $\xi^2$ 1 $\xi$ $\xi^2$ $\xi$ 1 |
| 15 | 022 112 202<br>1 $\xi^2$ $\xi^2$ $\xi$ $\xi$ $\xi^2$ $\xi^2$ 1 $\xi^2$ | P(2)<br>W(2) | $P_{01}(2)$<br>$Z(2)P_{12}(2)$ | 201 000 210<br>$\xi^2$ 1 $\xi$ 1 1 1 $\xi^2$ $\xi$ 1 |
| 16 | 022 121 220<br>1 $\xi^2$ $\xi^2$ $\xi$ $\xi^2$ $\xi$ $\xi^2$ $\xi^2$ 1 | P(2)<br>W(2) | $P_{01}(1) \otimes X^T(1)$<br>$Z^*(1)P_{12}(1) \otimes Z^*(1)$ | 210 000 201<br>$\xi^2$ $\xi$ 1 1 1 1 $\xi^2$ 1 $\xi$ |
| 17 | 022 202 112<br>1 $\xi^2$ $\xi^2$ $\xi^2$ 1 $\xi^2$ $\xi$ $\xi$ $\xi^2$ | P(2)<br>W(2) | $N(1) \otimes X^T(1)$<br>$Z^*(2)\cdot(P_{12}(1) \otimes I(1))$ | 201 210 000<br>$\xi^2$ 1 $\xi$ $\xi^2$ $\xi$ 1 1 1 1 |
| 18 | 022 220 121<br>1 $\xi^2$ $\xi^2$ $\xi^2$ $\xi^2$ 1 $\xi$ $\xi^2$ $\xi$ | P(2)<br>W(2) | $N(1) \otimes P_{01}(1)$<br>$(Z^*(1) \otimes Z(1))\cdot P_{12}(2)$ | 210 201 000<br>$\xi^2$ $\xi$ 1 $\xi^2$ 1 $\xi$ 1 1 1 |

Class 3: Reference function $f = [2\ 1\ 0\ 0\ 0\ 0\ 1\ 2]^T = x_1 x_2 \oplus (x_1)^2 \oplus 2x_2 \oplus 2 \mod 3$;

$F = [\xi^2\ \xi\ 1\ 1\ 1\ 1\ \xi\ \xi^2]^T$; $S_f = \mathbf{3} \cdot [1\ 1\ \xi^2\ \xi^2\ \xi^2\ \xi\ \xi^2\ \xi^2]^T$; exponents of $S_f$ = [002 212 122]

|  | $g$<br>$G = W(2) \cdot F$ |  |  | $[\![ S_g = P(2) S_f ]\!]$ |
|---|---|---|---|---|
| 2 | 210 012 000<br>$\xi^2\ \xi\ 1\ 1\ \xi\ \xi^2\ 1\ 1$ | P(2)<br>W(2) | $P_{12}(1) \otimes I(1)$<br>$P_{12}(1) \otimes I(1)$ | 002 122 212<br>$1\ 1\ \xi^2\ \xi\ \xi^2\ \xi^2\ \xi\ \xi^2$ |
| 3 | 210 111 201<br>$\xi^2\ \xi\ 1\ \xi\ \xi\ \xi^2\ 1\ \xi$ | P(2)<br>W(2) | $X(1) \otimes I(1)$<br>$Z(1) \otimes I(1)$ | 122 002 212<br>$\xi\ \xi^2\ \xi^2\ 1\ 1\ \xi^2\ \xi^2\ \xi\ \xi^2$ |
| 4 | 210 120 222<br>$\xi^2\ \xi\ 1\ \xi\ \xi^2\ 1\ \xi^2\ \xi^2\ \xi^2$ | P(2)<br>W(2) | $P_{01}(1) \otimes I(1)$<br>$Z(1)P_{12}(1) \otimes I(1)$ | 212 002 122<br>$\xi^2\ \xi\ \xi^2\ 1\ 1\ \xi^2\ \xi\ \xi^2\ \xi^2$ |
| 5 | 210 201 111<br>$\xi^2\ \xi\ 1\ \xi^2\ 1\ \xi\ \xi\ \xi$ | P(2)<br>W(2) | $N(1) \otimes I(1)$<br>$Z^*(1)P_{12}(1) \otimes I(1)$ | 122 212 002<br>$\xi\ \xi^2\ \xi^2\ \xi^2\ \xi\ \xi^2\ 1\ 1\ \xi^2$ |
| 6 | 210 222 120<br>$\xi^2\ \xi\ 1\ \xi^2\ \xi^2\ \xi^2\ \xi\ \xi^2\ 1$ | P(2)<br>W(2) | $X^T(1) \otimes I(1)$<br>$Z^*(1) \otimes I(1)$ | 212 122 002<br>$\xi^2\ \xi\ \xi^2\ \xi^2\ \xi^2\ \xi^2\ 1\ 1\ \xi^2$ |
| 7 | 222 012 021<br>$\xi^2\ \xi^2\ \xi^2\ 1\ \xi\ \xi^2\ 1\ \xi^2\ \xi$ | P(2)<br>W(2) | $I(1) \otimes X(1)$<br>$I(1) \otimes Z(1)$ | 200 221 212<br>$\xi^2\ 1\ 1\ \xi^2\ \xi^2\ \xi\ \xi^2\ \xi\ \xi^2$ |
| 8 | 201 000 021<br>$\xi^2\ 1\ \xi\ 1\ 1\ 1\ \xi^2\ \xi$ | P(2)<br>W(2) | $I(1) \otimes P_{12}(1)$<br>$I(1) \otimes P_{12}(1)$ | 020 221 122<br>$1\ \xi^2\ 1\ \xi^2\ \xi^2\ \xi\ \xi\ \xi^2\ \xi^2$ |
| 9 | 222 102 201<br>$\xi^2\ \xi^2\ \xi^2\ \xi\ 1\ \xi^2\ \xi^2\ 1\ \xi$ | P(2)<br>W(2) | $P_{01}(1) \otimes X(1)$<br>$Z(1) \cdot P_{12}(1) \otimes Z(1)$ | 221 200 212<br>$\xi^2\ \xi^2\ \xi\ \xi^2\ 1\ 1\ \xi^2\ \xi\ \xi^2$ |
| 10 | 201 102 222<br>$\xi^2\ 1\ \xi\ \xi\ 1\ \xi^2\ \xi^2\ \xi^2\ \xi^2$ | P(2)<br>W(2) | $P_{01}(1) \otimes P_{12}(1)$<br>$(Z(1) \otimes I(1)) \cdot P_{12}(2)$ | 221 020 122<br>$\xi^2\ \xi^2\ \xi\ 1\ \xi^2\ 1\ \xi\ \xi^2\ \xi^2$ |
| 11 | 201 210 111<br>$\xi^2\ 1\ \xi\ \xi^2\ \xi\ 1\ \xi\ \xi\ \xi$ | P(2)<br>W(2) | $N(1) \otimes P_{12}(1)$<br>$(Z^*(1) \otimes I(1)) P_{12}(2)$ | 122 221 020<br>$\xi\ \xi^2\ \xi^2\ \xi^2\ \xi^2\ \xi\ 1\ \xi^2\ 1$ |
| 12 | 222 210 120<br>$\xi^2\ \xi^2\ \xi^2\ \xi^2\ \xi\ 1\ \xi\ \xi^2\ 1$ | P(2)<br>W(2) | $N(1) \otimes X(1)$<br>$Z(1)^*P_{12}(1) \otimes Z(1)$ | 212 221 200<br>$\xi^2\ \xi\ \xi^2\ \xi^2\ \xi^2\ \xi\ \xi^2\ 1\ 1$ |
| 13 | 222 120 210<br>$\xi^2\ \xi^2\ \xi^2\ \xi\ \xi^2\ 1\ \xi^2\ \xi\ 1$ | P(2)<br>W(2) | $P_{01}(1) \otimes N(1)$<br>$Z(1) \cdot P_{12}(1) \otimes Z(1)^*P_{12}(1)$ | 212 200 221<br>$\xi^2\ \xi\ \xi^2\ \xi^2\ 1\ 1\ \xi^2\ \xi^2\ \xi$ |
| 14 | 201 021 000<br>$\xi^2\ 1\ \xi\ 1\ \xi^2\ \xi\ 1\ 1\ 1$ | P(2)<br>W(2) | $I(1) \otimes X^T(1)$<br>$I(1) \otimes Z^*(1)$ | 020 122 221<br>$1\ \xi^2\ 1\ \xi\ \xi^2\ \xi^2\ \xi^2\ \xi^2\ \xi$ |
| 15 | 222 021 012<br>$\xi^2\ \xi^2\ \xi^2\ 1\ \xi^2\ \xi\ 1\ \xi\ \xi^2$ | P(2)<br>W(2) | $P_{12}(1) \otimes X(1)$<br>$P_{12}(1) \otimes Z(1)$ | 200 212 221<br>$\xi^2\ 1\ 1\ \xi^2\ \xi\ \xi^2\ \xi^2\ \xi^2\ \xi$ |
| 16 | 201 111 210<br>$\xi^2\ 1\ \xi\ \xi\ \xi\ \xi\ \xi^2\ \xi\ 1$ | P(2)<br>W(2) | $P_{01}(1) \otimes X^T(1)$<br>$Z^*(1)P_{12}(1) \otimes Z^*(1)$ | 122 020 221<br>$\xi\ \xi^2\ \xi^2\ 1\ \xi^2\ 1\ \xi^2\ \xi^2\ \xi$ |
| 17 | 222 201 102<br>$\xi^2\ \xi^2\ \xi^2\ \xi^2\ 1\ \xi\ \xi\ 1\ \xi^2$ | P(2)<br>W(2) | $X^T(1) \otimes X(1)$<br>$Z^*(1) \otimes Z(1)$ | 221 212 200<br>$\xi^2\ \xi^2\ \xi\ \xi^2\ \xi\ \xi^2\ \xi^2\ 1\ 1$ |
| 18 | 201 222 102<br>$\xi^2\ 1\ \xi\ \xi^2\ \xi^2\ \xi^2\ \xi\ 1\ \xi^2$ | P(2)<br>W(2) | $N(1) \otimes X^T(1)$<br>$Z^*(2) \cdot (P_{12}(1) \otimes I(1))$ | 221 122 020<br>$\xi^2\ \xi^2\ \xi\ \xi\ \xi^2\ \xi^2\ 1\ \xi^2\ 1$ |

Class 4: Reference function $f = [1\ 0\ 0\ 0\ 1\ 0\ 2\ 2\ 0]^T = 2x_1x_2 \oplus 2x_1 \oplus 2(x_2)^2 \oplus 1 \mod 3$;

$F = [\ \xi\ 1\ 1\ 1\ \xi\ 1\ \xi^2\ \xi^2\ 1\ ]^T$ ; $S_f = 3[\ 1\ \xi\ \xi^2\ 1\ \xi^2\ \xi\ \xi\ \xi\ \xi\ ]^T$; exponents of $S_f$ = [012 021 111]

| | g<br>G = W(2)·F | | | $[\![S_g = P(2)S_f]\!]$ |
|---|---|---|---|---|
| 2 | 100 001 202<br>$\xi\ 1\ 1\ 1\ 1\ \xi\ \xi^2\ 1\ \xi^2$ | P(2)<br>W(2) | $I(1) \otimes P_{12}(1)$<br>$I(1) \otimes P_{12}(1)$ | 021 012 111<br>$1\ \xi^2\ \xi\ 1\ \xi^2\ \xi\ \xi\ \xi$ |
| 3 | 100 112 121<br>$\xi\ 1\ 1\ \xi\ \xi\ \xi^2\ \xi\ \xi^2\ \xi$ | P(2)<br>W(2) | $N(1) \otimes I(1)$<br>$Z^*(1)P_{12}(1) \otimes I(1)$ | 111 021 012<br>$\xi\ \xi\ \xi\ 1\ \xi^2\ \xi\ 1\ \xi\ \xi^2$ |
| 4 | 100 121 112<br>$\xi\ 1\ 1\ \xi\ \xi^2\ \xi\ \xi\ \xi\ \xi^2$ | P(2)<br>W(2) | $N(1) \otimes P_{12}(1)$<br>$Z(1)^*P_{12}(1) \otimes P_{12}(1)$ | 111 012 021<br>$\xi\ \xi\ \xi\ 1\ \xi\ \xi^2\ 1\ \xi^2\ \xi$ |
| 5 | 100 202 001<br>$\xi\ 1\ 1\ \xi^2\ 1\ \xi^2\ 1\ 1\ \xi$ | P(2)<br>W(2) | $P_{12}(2)$<br>$P_{12}(2)$ | 021 111 012<br>$1\ \xi^2\ \xi\ \xi\ \xi\ \xi\ 1\ \xi\ \xi^2$ |
| 6 | 100 220 010<br>$\xi\ 1\ 1\ \xi^2\ \xi^2\ 1\ 1\ \xi\ 1$ | P(2)<br>W(2) | $X^T(1) \otimes P_{12}(1)$<br>$Z^*(1) \otimes P_{12}(1)$ | 012 111 021<br>$1\ \xi\ \xi^2\ \xi\ \xi\ \xi\ 1\ \xi^2\ \xi$ |
| 7 | 112 010 211<br>$\xi\ \xi\ \xi^2\ 1\ \xi\ 1\ \xi^2\ \xi\ \xi$ | P(2)<br>W(2) | $P_{01}(1) \otimes X(1)$<br>$Z(1) \cdot P_{12}(2) \otimes Z(1)$ | 102 201 111<br>$\xi\ 1\ \xi^2\ \xi^2\ 1\ \xi\ \xi\ \xi\ \xi$ |
| 8 | 112 022 202<br>$\xi\ \xi\ \xi^2\ 1\ \xi^2\ \xi^2\ \xi^2\ 1\ \xi^2$ | P(2)<br>W(2) | $I(1) \otimes X(1)$<br>$I(1) \otimes Z(1)$ | 201 102 111<br>$\xi^2\ 1\ \xi\ \xi\ 1\ \xi^2\ \xi\ \xi\ \xi$ |
| 9 | 112 100 121<br>$\xi\ \xi\ \xi^2\ \xi\ 1\ 1\ \xi\ \xi^2\ \xi$ | P(2)<br>W(2) | $N(1) \otimes P_{01}(1)$<br>$(Z^*(1) \otimes Z(1)) \cdot P_{12}(2)$ | 111 201 102<br>$\xi\ \xi\ \xi\ \xi^2\ 1\ \xi\ \xi\ 1\ \xi^2$ |
| 10 | 112 121 100<br>$\xi\ \xi\ \xi^2\ \xi\ \xi^2\ \xi\ \xi\ 1\ 1$ | P(2)<br>W(2) | $N(1) \otimes X(1)$<br>$Z(1)^*P_{12}(1) \otimes Z(1)$ | 111 102 201<br>$\xi\ \xi\ \xi\ \xi\ 1\ \xi^2\ \xi^2\ 1\ \xi$ |
| 11 | 112 202 022<br>$\xi\ \xi\ \xi^2\ \xi^2\ 1\ \xi^2\ 1\ \xi^2\ \xi^2$ | P(2)<br>W(2) | $X^T(1) \otimes P_{01}(1)$<br>$Z^*(1) \otimes Z^*(1)P_{12}(1)$ | 201 111 102<br>$\xi^2\ 1\ \xi\ \xi\ \xi\ \xi\ \xi\ 1\ \xi^2$ |
| 12 | 112 211 010<br>$\xi\ \xi\ \xi^2\ \xi^2\ \xi\ \xi\ 1\ \xi\ 1$ | P(2)<br>W(2) | $X^T(1) \otimes X(1)$<br>$Z^*(1) \otimes Z(1)$ | 102 111 201<br>$\xi\ 1\ \xi^2\ \xi\ \xi\ \xi\ \xi^2\ 1\ \xi$ |
| 13 | 121 001 211<br>$\xi\ \xi^2\ \xi\ 1\ 1\ \xi\ \xi^2\ \xi\ \xi$ | P(2)<br>W(2) | $I(1) \otimes X^T(1)$<br>$I(1) \otimes Z^*(1)$ | 120 210 111<br>$\xi\ \xi^2\ 1\ \xi^2\ \xi\ 1\ \xi\ \xi\ \xi$ |
| 14 | 121 022 220<br>$\xi\ \xi^2\ \xi\ 1\ \xi^2\ \xi^2\ \xi^2\ \xi^2\ 1$ | P(2)<br>W(2) | $P_{01}(1) \otimes X^T(1)$<br>$Z^*(1)P_{12}(1) \otimes Z^*(1)$ | 210 120 111<br>$\xi^2\ \xi\ 1\ \xi\ \xi^2\ 1\ \xi\ \xi\ \xi$ |
| 15 | 121 100 112<br>$\xi\ \xi^2\ \xi\ \xi\ 1\ 1\ \xi\ \xi\ \xi^2$ | P(2)<br>W(2) | $N(1) \otimes X^T(1)$<br>$Z^*(2) \cdot (P_{12}(1) \otimes I(1))$ | 111 210 120<br>$\xi\ \xi\ \xi\ \xi^2\ \xi\ 1\ \xi\ \xi^2\ 1$ |
| 16 | 121 112 100<br>$\xi\ \xi^2\ \xi\ \xi\ \xi\ \xi^2\ \xi\ 1\ 1$ | P(2)<br>W(2) | $N(2)$<br>$Z^*(2) \cdot P_{12}(2)$ | 111 120 210<br>$\xi\ \xi\ \xi\ \xi\ \xi^2\ 1\ \xi^2\ \xi\ 1$ |
| 17 | 121 211 001<br>$\xi\ \xi^2\ \xi\ \xi^2\ \xi\ \xi\ 1\ 1\ \xi$ | P(2)<br>W(2) | $X^T(1) \otimes N(1)$<br>$Z^*(1) \otimes Z(1)^*P_{12}(1)$ | 120 111 210<br>$\xi\ \xi^2\ 1\ \xi\ \xi\ \xi\ \xi^2\ \xi\ 1$ |
| 18 | 121 220 022<br>$\xi\ \xi^2\ \xi\ \xi^2\ \xi^2\ 1\ 1\ \xi^2\ \xi^2$ | P(2)<br>W(2) | $P_{12}(1) \otimes N(1)$<br>$P_{12}(1) \otimes Z(1)^*P_{12}(1)$ | 210 111 120<br>$\xi^2\ \xi\ 1\ \xi\ \xi\ \xi\ \xi\ \xi^2\ 1$ |

Class 5. Reference function  $f = [2\ 0\ 0\ 1\ 1\ 0\ 0\ 2\ 0]^T = 2x_1x_2 \oplus 2x_1 \oplus (x_2)^2 \oplus 2\ mod\ 3$;

$F = [\ \xi^2\ 1\ 1\ \xi\ \xi\ 1\ 1\ \xi^2\ 1\ ]^T$ ; $S_f = 3\cdot[\ 1\ \xi\ \xi^2\ 1\ \xi^2\ \xi\ \xi^2\ \xi^2\ \xi^2\ ]^T$ ; exponents of $S_f$ = [012 021 222]

|  | $g$<br>G = W(2)·F |  |  | $[\![ S_g = P(2)S_f ]\!]$ |
|---|---|---|---|---|
| 2 | 200 002 101<br>$\xi^2\ 1\ 1\ 1\ 1\ \xi^2\ \xi\ 1\ \xi$ | P(2)<br>W(2) | $X^T(1) \otimes I(1)$<br>$Z^*(1) \otimes I(1)$ | 021 222 012<br>$1\ \xi^2\ \xi\ \xi^2\ \xi^2\ \xi^2\ 1\ \xi\ \xi^2$ |
| 3 | 200 020 110<br>$\xi^2\ 1\ 1\ 1\ \xi^2\ 1\ \xi\ \xi\ 1$ | P(2)<br>W(2) | $P_{12}(1) \otimes I(1)$<br>$P_{12}(1) \otimes I(1)$ | 012 222 021<br>$1\ \xi\ \xi^2\ \xi^2\ \xi^2\ \xi^2\ 1\ \xi^2\ \xi$ |
| 4 | 200 101 002<br>$\xi^2\ 1\ 1\ \xi\ 1\ \xi\ 1\ 1\ \xi^2$ | P(2)<br>W(2) | $P_{01}(1) \otimes I(1)$<br>$Z(1)P_{12}(1) \otimes I(1)$ | 021 012 222<br>$1\ \xi^2\ \xi\ 1\ \xi\ \xi^2\ \xi^2\ \xi^2\ \xi^2$ |
| 5 | 200 212 221<br>$\xi^2\ 1\ 1\ \xi^2\ \xi\ \xi^2\ \xi^2\ \xi^2\ \xi$ | P(2)<br>W(2) | $N(1) \otimes I(1)$<br>$Z^*(1)P_{12}(1) \otimes I(1)$ | 222 021 012<br>$\xi^2\ \xi^2\ \xi^2\ 1\ \xi^2\ \xi\ 1\ \xi\ \xi^2$ |
| 6 | 200 221 212<br>$\xi^2\ 1\ 1\ \xi^2\ \xi^2\ \xi\ \xi^2\ \xi^2\ \xi^2$ | P(2)<br>W(2) | $X(1) \otimes I(1)$<br>$Z(1) \otimes I(1)$ | 222 012 021<br>$\xi^2\ \xi^2\ \xi^2\ 1\ \xi\ \xi^2\ 1\ \xi^2\ \xi$ |
| 7 | 212 002 122<br>$\xi^2\ \xi\ \xi^2\ 1\ 1\ \xi^2\ \xi\ \xi^2\ \xi^2$ | P(2)<br>W(2) | $X^T(1) \otimes P_{01}(1)$<br>$Z^*(1) \otimes P_{12}(1)$ | 201 222 102<br>$\xi^2\ 1\ \xi\ \xi^2\ \xi^2\ \xi^2\ \xi\ 1\ \xi^2$ |
| 8 | 212 011 110<br>$\xi^2\ \xi\ \xi^2\ 1\ \xi\ \xi\ \xi\ \xi\ 1$ | P(2)<br>W(2) | $X^T(1) \otimes X(1)$<br>$Z^*(1) \otimes Z(1)$ | 102 222 201<br>$\xi\ 1\ \xi^2\ \xi^2\ \xi^2\ \xi^2\ \xi^2\ 1\ \xi$ |
| 9 | 212 110 011<br>$\xi^2\ \xi\ \xi^2\ \xi\ \xi\ 1\ 1\ \xi\ \xi$ | P(2)<br>W(2) | $I(1) \otimes P_{01}(1)$<br>$I(1) \otimes Z(1)P_{12}(1)$ | 102 201 222<br>$\xi\ 1\ \xi^2\ \xi^2\ 1\ \xi\ \xi^2\ \xi^2\ \xi^2$ |
| 10 | 212 122 002<br>$\xi^2\ \xi\ \xi^2\ \xi\ \xi^2\ \xi^2\ 1\ 1\ \xi^2$ | P(2)<br>W(2) | $I(1) \otimes X(1)$<br>$I(1) \otimes Z(1)$ | 201 102 222<br>$\xi^2\ 1\ \xi\ \xi\ 1\ \xi^2\ \xi^2\ \xi^2\ \xi^2$ |
| 11 | 212 200 221<br>$\xi^2\ \xi\ \xi^2\ \xi^2\ 1\ 1\ \xi^2\ \xi^2\ \xi$ | P(2)<br>W(2) | $N(1) \otimes P_{01}(1)$<br>$(Z^*(1) \otimes Z(1))\cdot P_{12}(2)$ | 222 201 102<br>$\xi^2\ \xi^2\ \xi^2\ \xi^2\ 1\ \xi\ \xi\ 1\ \xi^2$ |
| 12 | 212 221 200<br>$\xi^2\ \xi\ \xi^2\ \xi^2\ \xi^2\ \xi\ \xi^2\ 1\ 1$ | P(2)<br>W(2) | $N(1) \otimes X(1)$<br>$Z^*(1)P_{12}(1) \otimes Z(1)$ | 222 102 201<br>$\xi^2\ \xi^2\ \xi^2\ \xi\ 1\ \xi^2\ \xi^2\ 1\ \xi$ |
| 13 | 221 011 101<br>$\xi^2\ \xi^2\ \xi\ 1\ \xi\ \xi\ \xi\ 1\ \xi$ | P(2)<br>W(2) | $X^T(1) \otimes N(1)$<br>$Z^*(1) \otimes Z^*(1)P_{12}(1)$ | 120 222 210<br>$\xi\ \xi^2\ 1\ \xi^2\ \xi^2\ \xi^2\ \xi^2\ \xi\ 1$ |
| 14 | 221 020 122<br>$\xi^2\ \xi^2\ \xi\ 1\ \xi^2\ 1\ \xi\ \xi^2\ \xi^2$ | P(2)<br>W(2) | $P_{12}(1) \otimes N(1)$<br>$P_{12}(1) \otimes Z^*(1)P_{12}(1)$ | 210 222 120<br>$\xi^2\ \xi\ 1\ \xi^2\ \xi^2\ \xi^2\ \xi\ \xi^2\ 1$ |
| 15 | 221 101 011<br>$\xi^2\ \xi^2\ \xi\ \xi\ 1\ \xi\ 1\ \xi\ \xi$ | P(2)<br>W(2) | $I(1) \otimes X^T(1)$<br>$I(1) \otimes Z^*(1)$ | 120 210 222<br>$\xi\ \xi^2\ 1\ \xi^2\ \xi\ 1\ \xi^2\ \xi^2\ \xi^2$ |
| 16 | 221 122 020<br>$\xi^2\ \xi^2\ \xi\ \xi\ \xi^2\ \xi^2\ 1\ \xi^2\ 1$ | P(2)<br>W(2) | $I(1) \otimes N(1)$<br>$I(1) \otimes Z^*(1)P_{12}(1)$ | 210 120 222<br>$\xi^2\ \xi\ 1\ \xi\ \xi^2\ 1\ \xi^2\ \xi^2\ \xi^2$ |
| 17 | 221 200 212<br>$\xi^2\ \xi^2\ \xi\ \xi^2\ 1\ 1\ \xi^2\ \xi\ \xi^2$ | P(2)<br>W(2) | $N(1) \otimes X^T(1)$<br>$Z^*(1)P_{12}(1) \otimes Z^*(1)$ | 222 210 120<br>$\xi^2\ \xi^2\ \xi^2\ \xi^2\ \xi\ 1\ \xi\ \xi^2\ 1$ |
| 18 | 221 212 200<br>$\xi^2\ \xi^2\ \xi\ \xi^2\ \xi\ \xi\ \xi^2\ 1\ 1$ | P(2)<br>W(2) | $N(2)$<br>$Z^*(2)\cdot P_{12}(2)$ | 222 120 210<br>$\xi^2\ \xi^2\ \xi^2\ \xi\ \xi^2\ 1\ \xi^2\ \xi\ 1$ |

Class 6: Reference function $f = [1\ 0\ 2\ 0\ 0\ 0\ 0\ 1\ 2]^T = x_1 x_2 \oplus 2x_2 \oplus 2(x_1)^2 \oplus 1\ mod\ 3$;

$F = [\xi\ 1\ \xi^2\ 1\ 1\ 1\ 1\ \xi\ \xi^2]^T$ ; $S_f = 3\cdot[1\ 1\ \xi\ \xi^2\ \xi\ \xi\ \xi\ \xi^2\ \xi]^T$ : exponents of $S_f = [001\ 211\ 121]$

| | $g$<br>$G = W(2)\cdot F$ | | | $[\![S_g = P(2)S_f]\!]$ |
|---|---|---|---|---|
| 2 | 102 012 000<br>$\xi\ 1\ \xi^2\ 1\ \xi\ \xi^2\ 1\ 1\ 1$ | P(2)<br>W(2) | $P_{12}(1) \otimes I(1)$<br>$P_{12}(1) \otimes I(1)$ | 001 121 211<br>$1\ 1\ \xi\ \xi\ \xi^2\ \xi\ \xi^2\ \xi\ \xi$ |
| 3 | 102 111 201<br>$\xi\ 1\ \xi^2\ \xi\ \xi\ \xi^2\ 1\ \xi$ | P(2)<br>W(2) | $P_{01}(2)$<br>$Z(2)\cdot P_{12}(2)$ | 121 001 211<br>$\xi\ \xi^2\ \xi\ 1\ 1\ \xi\ \xi^2\ \xi\ \xi$ |
| 4 | 102 120 222<br>$\xi\ 1\ \xi^2\ \xi\ \xi^2\ 1\ \xi^2\ \xi^2\ \xi^2$ | P(2)<br>W(2) | $X(1) \otimes P_{01}(1)$<br>$Z(1) \otimes Z(1)P_{12}(1)$ | 211 001 121<br>$\xi^2\ \xi\ \xi\ 1\ 1\ \xi\ \xi\ \xi^2\ \xi$ |
| 5 | 102 201 111<br>$\xi\ 1\ \xi^2\ \xi^2\ 1\ \xi\ \xi\ \xi\ \xi$ | P(2)<br>W(2) | $X^T(1) \otimes P_{01}(1)$<br>$Z^*(1) \otimes Z(1)P_{12}(1)$ | 121 211 001<br>$\xi\ \xi^2\ \xi\ \xi^2\ \xi\ \xi\ 1\ 1\ \xi$ |
| 6 | 102 222 120<br>$\xi\ 1\ \xi^2\ \xi^2\ \xi^2\ \xi^2\ \xi\ \xi^2\ 1$ | P(2)<br>W(2) | $N(1) \otimes P_{01}(1)$<br>$(Z^*(1) \otimes Z(1))\cdot P_{12}(2)$ | 211 121 001<br>$\xi^2\ \xi\ \xi\ \xi^2\ \xi\ 1\ 1\ \xi$ |
| 7 | 111 102 201<br>$\xi\ \xi\ \xi\ 1\ \xi^2\ \xi^2\ 1\ \xi$ | P(2)<br>W(2) | $P_{01}(1) \otimes X(1)$<br>$Z(1)\cdot P_{12}(1) \otimes Z(1)$ | 121 100 112<br>$\xi\ \xi^2\ \xi\ \xi\ 1\ 1\ \xi\ \xi\ \xi^2$ |
| 8 | 111 012 021<br>$\xi\ \xi\ \xi\ 1\ \xi\ \xi^2\ 1\ \xi^2\ \xi$ | P(2)<br>W(2) | $P_{12}(1) \otimes N(1)$<br>$P_{12}(1) \otimes Z^*(1)P_{12}(1)$ | 100 121 112<br>$\xi\ 1\ 1\ \xi\ \xi^2\ \xi\ \xi\ \xi\ \xi^2$ |
| 9 | 111 021 012<br>$\xi\ \xi\ \xi\ 1\ \xi^2\ \xi\ 1\ \xi^2$ | P(2)<br>W(2) | $P_{12}(1) \otimes X(1)$<br>$P_{12}(1) \otimes Z(1)$ | 100 112 121<br>$\xi\ 1\ 1\ \xi\ \xi\ \xi^2\ \xi\ \xi^2\ \xi$ |
| 10 | 111 120 210<br>$\xi\ \xi\ \xi\ \xi\ \xi^2\ 1\ \xi^2\ \xi\ 1$ | P(2)<br>W(2) | $P_{01}(1) \otimes N(1)$<br>$Z(1)\cdot P_{12}(1) \otimes Z^*(1)P_{12}(1)$ | 112 100 121<br>$\xi\ \xi\ \xi^2\ \xi\ 1\ 1\ \xi\ \xi^2\ \xi$ |
| 11 | 111 201 102<br>$\xi\ \xi\ \xi\ \xi^2\ 1\ \xi\ \xi\ 1\ \xi^2$ | P(2)<br>W(2) | $N(2)$<br>$Z^*(2)\cdot P_{12}(2)$ | 121 112 100<br>$\xi\ \xi^2\ \xi\ \xi\ \xi\ \xi^2\ \xi\ 1\ 1$ |
| 12 | 111 210 120<br>$\xi\ \xi\ \xi\ \xi^2\ \xi\ 1\ \xi\ \xi^2\ 1$ | P(2)<br>W(2) | $X^T(1) \otimes N(1)$<br>$Z^*(1) \otimes Z^*(1)P12(1)$ | 112 121 100<br>$\xi\ \xi\ \xi^2\ \xi\ \xi^2\ \xi\ \xi\ 1\ 1$ |
| 13 | 120 000 021<br>$\xi\ \xi^2\ 1\ 1\ 1\ 1\ 1\ \xi^2\ \xi$ | P(2)<br>W(2) | $P_{12}(1) \otimes X^T(1)$<br>$P_{12}(1) \otimes Z^*(1)$ | 010 211 112<br>$1\ \xi\ 1\ \xi^2\ \xi\ \xi\ \xi\ \xi^2$ |
| 14 | 120 021 000<br>$\xi\ \xi^2\ 1\ 1\ \xi^2\ \xi\ 1\ 1\ 1$ | P(2)<br>W(2) | $P_{12}(2)$<br>$P_{12}(2)$ | 010 112 211<br>$1\ \xi\ 1\ \xi\ \xi\ \xi^2\ \xi^2\ \xi\ \xi$ |
| 15 | 120 102 222<br>$\xi\ \xi^2\ 1\ \xi\ 1\ \xi^2\ \xi^2\ \xi^2\ \xi^2$ | P(2)<br>W(2) | $P_{01}(1) \otimes P_{12}(1)$<br>$Z(1)\cdot P_{12}(1) \otimes P_{12}(1)$ | 211 010 112<br>$\xi^2\ \xi\ \xi\ 1\ \xi\ 1\ \xi\ \xi\ \xi^2$ |
| 16 | 120 111 210<br>$\xi\ \xi^2\ 1\ \xi\ \xi\ \xi\ \xi^2\ \xi\ 1$ | P(2)<br>W(2) | $P_{01}(1) \otimes X^T(1)$<br>$Z(1)\cdot P_{12}(1) \otimes Z^*(1)$ | 112 010 211<br>$\xi\ \xi\ \xi^2\ 1\ \xi\ 1\ \xi^2\ \xi\ \xi$ |
| 17 | 120 210 111<br>$\xi\ \xi^2\ 1\ \xi^2\ \xi\ 1\ \xi\ \xi\ \xi$ | P(2)<br>W(2) | $X^T(1) \otimes X^T(1)$<br>$Z^*(1) \otimes Z^*(1)$ | 112 211 010<br>$\xi\ \xi\ \xi^2\ \xi^2\ \xi\ \xi\ 1\ \xi\ 1$ |
| 18 | 120 222 102<br>$\xi\ \xi^2\ 1\ \xi^2\ \xi^2\ \xi^2\ \xi\ 1\ \xi^2$ | P(2)<br>W(2) | $N(1) \otimes X^T(1)$<br>$Z^*(1)P_{12}(1) \otimes Z^*(1)$ | 211 112 010<br>$\xi^2\ \xi\ \xi\ \xi\ \xi\ \xi^2\ 1\ \xi\ 1$ |

Class 7: Reference function $f = [\,0\,0\,0\,2\,0\,1\,0\,2\,1\,]^T = x_1x_2 \oplus x_1 \oplus (x_1)^2 \ mod\ 3$;

$F = [\,1\,1\,1\,\xi^2\,1\,\xi\,1\,\xi^2\,\xi\,]^T$ ; $S_f = 3\,[\,1\,\xi^2\,1\,1\,\xi\,\xi\,1\,1\,\xi^2\,]^T$; exponents of $S_f$ = [020 011 002]

| | $g$<br>$G = W(2)\cdot F$ | | | $[\![S_g = P(2)S_f]\!]$ |
|---|---|---|---|---|
| 2 | 000 210 012<br>$1\,1\,1\,\xi^2\,1\,1\,\xi\,\xi^2$ | P(2)<br>W(2) | $N(1) \otimes I(1)$<br>$Z^*(1)P_{12}(1) \otimes I(1)$ | 002 011 020<br>$1\,1\,\xi^2\,1\,\xi\,\xi\,1\,\xi^2\,1$ |
| 3 | 000 012 210<br>$1\,1\,1\,1\,\xi\,\xi^2\,\xi^2\,\xi\,1$ | P(2)<br>W(2) | $X(1) \otimes I(1)$<br>$Z(1) \otimes I(1)$ | 002 020 011<br>$1\,1\,\xi^2\,1\,\xi^2\,1\,1\,\xi\,\xi$ |
| 4 | 012 210 000<br>$1\,\xi\,\xi^2\,\xi^2\,\xi\,1\,1\,1\,1$ | P(2)<br>W(2) | $I(1) \otimes X(1)$<br>$I(1) \otimes Z(1)$ | 002 101 200<br>$1\,1\,\xi^2\,\xi\,1\,\xi\,\xi^2\,1\,1$ |
| 5 | 012 000 210<br>$1\,\xi\,\xi^2\,1\,1\,1\,\xi^2\,\xi\,1$ | P(2)<br>W(2) | $P_{12}(1) \otimes X(1)$<br>$P_{12}(1) \otimes Z(1)$ | 002 200 101<br>$1\,1\,\xi^2\,\xi^2\,1\,1\,\xi\,1\,\xi$ |
| 6 | 000 120 102<br>$1\,1\,1\,\xi\,\xi^2\,1\,\xi\,1\,\xi^2$ | P(2)<br>W(2) | $X^T(1) \otimes I(1)$<br>$Z^*(1) \otimes I(1)$ | 011 002 020<br>$1\,\xi\,\xi\,1\,1\,\xi^2\,1\,\xi^2\,1$ |
| 7 | 000 102 120<br>$1\,1\,1\,\xi\,1\,\xi^2\,\xi\,\xi^2\,1$ | P(2)<br>W(2) | $P_{01}(1) \otimes I(1)$<br>$Z(1)\cdot P_{12}(1) \otimes I(1)$ | 011 020 002<br>$1\,\xi\,\xi\,1\,\xi^2\,1\,1\,1\,\xi^2$ |
| 8 | 000 021 201<br>$1\,1\,1\,1\,\xi^2\,\xi\,\xi^2\,1\,\xi$ | P(2)<br>W(2) | $P_{12}(1) \otimes I(1)$<br>$P_{12}(1) \otimes I(1)$ | 020 002 011<br>$1\,\xi^2\,1\,1\,1\,\xi^2\,1\,\xi\,\xi$ |
| 9 | 021 201 000<br>$1\,\xi^2\,\xi\,\xi^2\,1\,\xi\,1\,1\,1$ | P(2)<br>W(2) | $I(1) \otimes N(1)$<br>$I(1) \otimes Z^*(1)P_{12}(1)$ | 020 110 200<br>$1\,\xi^2\,1\,\xi\,\xi\,1\,\xi^2\,1\,1$ |
| 10 | 021 000 201<br>$1\,\xi^2\,\xi\,1\,1\,1\,\xi^2\,1\,\xi$ | P(2)<br>W(2) | $P_{12}(1) \otimes N(1)$<br>$P_{12}(1) \otimes Z^*(1)P_{12}(1)$ | 020 200 110<br>$1\,\xi^2\,1\,\xi^2\,1\,1\,\xi\,\xi\,1$ |
| 11 | 012 111 102<br>$1\,\xi\,\xi^2\,\xi\,\xi\,\xi\,1\,\xi^2$ | P(2)<br>W(2) | $P_{01}(1) \otimes X(1)$<br>$Z(1)\cdot P_{12}(1) \otimes Z(1)$ | 101 002 200<br>$\xi\,1\,\xi\,1\,1\,\xi^2\,\xi^2\,1\,1$ |
| 12 | 012 102 111<br>$1\,\xi\,\xi^2\,\xi\,1\,\xi^2\,\xi\,\xi\,\xi$ | P(2)<br>W(2) | $P_{01}(2)$<br>$Z(2)\cdot P_{12}(2)$ | 101 200 002<br>$\xi\,1\,\xi\,\xi^2\,1\,1\,1\,1\,\xi^2$ |
| 13 | 021 111 120<br>$1\,\xi^2\,\xi\,\xi\,\xi\,\xi\,\xi\,\xi^2\,1$ | P(2)<br>W(2) | $P_{01}(1) \otimes N(1)$<br>$Z(1)\cdot P_{12}(1) \otimes Z^*(1)P_{12}(1)$ | 110 020 200<br>$\xi\,\xi\,1\,1\,\xi^2\,1\,\xi^2\,1\,1$ |
| 14 | 021 120 111<br>$1\,\xi^2\,\xi\,\xi\,\xi^2\,1\,\xi\,\xi\,\xi$ | P(2)<br>W(2) | $P_{01}(1) \otimes X^T(1)$<br>$Z^*(1)P_{12}(1) \otimes Z^*(1)$ | 110 200 020<br>$\xi\,\xi\,1\,\xi^2\,1\,1\,1\,\xi^2\,1$ |
| 15 | 012 021 222<br>$1\,\xi\,\xi^2\,1\,\xi^2\,\xi\,\xi^2\,\xi^2\,\xi^2$ | P(2)<br>W(2) | $P_{12}(1) \otimes P_{01}(1)$<br>$P_{12}(1) \otimes Z^*(1)P_{12}(1)$ | 200 002 101<br>$\xi^2\,1\,1\,1\,1\,\xi^2\,\xi\,1\,\xi$ |
| 16 | 021 012 222<br>$1\,\xi^2\,\xi\,1\,\xi\,\xi^2\,\xi^2\,\xi^2\,\xi^2$ | P(2)<br>W(2) | $P_{12}(1) \otimes X^T(1)$<br>$P_{12}(1) \otimes Z^*(1)$ | 200 020 110<br>$\xi^2\,1\,1\,1\,\xi^2\,1\,\xi\,\xi\,1$ |
| 17 | 012 222 021<br>$1\,\xi\,\xi^2\,\xi^2\,\xi^2\,\xi^2\,1\,\xi^2\,\xi$ | P(2)<br>W(2) | $I(1) \otimes P_{01}(1)$<br>$I(1) \otimes Z^*(1)P_{12}(1)$ | 200 101 002<br>$\xi^2\,1\,1\,\xi\,1\,\xi\,1\,1\,\xi^2$ |
| 18 | 021 222 012<br>$1\,\xi^2\,\xi\,\xi^2\,\xi^2\,\xi^2\,1\,\xi\,\xi^2$ | P(2)<br>W(2) | $I(1) \otimes X^T(1)$<br>$I(1) \otimes Z^*(1)$ | 200 110 020<br>$\xi^2\,1\,1\,\xi\,\xi\,1\,1\,\xi^2\,1$ |

Class 8: Reference function $f = [\,0\,0\,0\,0\,2\,1\,1\,2\,0\,]^T$; $= 2x_1x_2 \oplus x_1 \oplus 2(x_1)^2 \mod 3$;

$F = [\,1\,1\,1\,1\,\xi^2\,\xi\,\xi\,\xi^2\,1\,]^T$; $S_f = 3\,[\,1\,\xi\,1\,1\,\xi^2\,\xi^2\,1\,1\,\xi\,]^T$; exponents of $S_f$ = [ 010 022 001]

| | $g$<br>$G = W(2)\cdot F$ | | | $[\![S_g = P(2)S_f]\!]$ |
|---|---|---|---|---|
| 2 | 000 102 012<br>$1\,1\,1\,1\,\xi\,1\,\xi^2\,1\,\xi\,\xi^2$ | P(2)<br>W(2) | $X(1) \otimes I(1)$<br>$Z(1) \otimes I(1)$ | 001 010 022<br>$1\,1\,\xi\,1\,\xi\,1\,1\,\xi^2\,\xi^2$ |
| 3 | 000 012 102<br>$1\,1\,1\,1\,\xi\,\xi^2\,\xi\,1\,\xi^2$ | P(2)<br>W(2) | $N(1) \otimes I(1)$<br>$Z^*(1)P_{12}(1) \otimes I(1)$ | 001 022 010<br>$1\,1\,\xi\,1\,\xi^2\,\xi^2\,1\,\xi\,1$ |
| 4 | 012 102 000<br>$1\,\xi\,\xi^2\,\xi\,1\,\xi^2\,1\,1\,1$ | P(2)<br>W(2) | $P_{12}(1) \otimes X(1)$<br>$P_{12}(1) \otimes Z(1)$ | 001 100 202<br>$1\,1\,\xi\,\xi\,1\,1\,\xi^2\,1\,\xi^2$ |
| 5 | 012 000 102<br>$1\,\xi\,\xi^2\,1\,1\,1\,\xi\,1\,\xi^2$ | P(2)<br>W(2) | $I(1) \otimes X(1)$<br>$I(1) \otimes Z(1)$ | 001 202 100<br>$1\,1\,\xi\,\xi^2\,1\,\xi^2\,\xi\,1\,1$ |
| 6 | 000 210 201<br>$1\,1\,1\,\xi^2\,\xi\,1\,\xi^2\,1\,\xi$ | P(2)<br>W(2) | $X^T(1) \otimes I(1)$<br>$Z^*(1) \otimes I(1)$ | 022 001 010<br>$1\,\xi^2\,\xi^2\,1\,1\,\xi\,1\,\xi\,1$ |
| 7 | 000 201 210<br>$1\,1\,1\,\xi^2\,1\,\xi\,\xi^2\,\xi\,1$ | P(2)<br>W(2) | $P_{01}(1) \otimes I(1)$<br>$Z(1)P_{12}(1) \otimes I(1)$ | 022 010 001<br>$1\,\xi^2\,\xi^2\,1\,\xi\,1\,1\,1\,\xi$ |
| 8 | 000 120 021<br>$1\,1\,1\,\xi\,\xi^2\,1\,1\,\xi^2\,\xi$ | P(2)<br>W(2) | $P_{12}(1) \otimes I(1)$<br>$P_{12}(1) \otimes I(1)$ | 010 001 022<br>$1\,\xi\,1\,1\,1\,\xi\,1\,\xi^2\,\xi^2$ |
| 9 | 021 120 000<br>$1\,\xi^2\,\xi\,\xi\,\xi^2\,1\,1\,1\,1$ | P(2)<br>W(2) | $P_{12}(1) \otimes N(1)$<br>$P_{12}(1) \otimes Z^*(1)P_{12}(1)$ | 010 100 220<br>$1\,\xi\,1\,\xi\,1\,1\,\xi^2\,\xi^2\,1$ |
| 10 | 021 000 120<br>$1\,\xi^2\,\xi\,1\,1\,1\,\xi\,\xi^2\,1$ | P(2)<br>W(2) | $I(1) \otimes N(1)$<br>$I(1) \otimes Z^*(1)P_{12}(1)$ | 010 220 100<br>$1\,\xi\,1\,\xi^2\,\xi^2\,1\,\xi\,1\,1$ |
| 11 | 012 111 021<br>$1\,\xi\,\xi^2\,\xi\,\xi\,\xi\,1\,\xi^2\,\xi$ | P(2)<br>W(2) | $P_{12}(1) \otimes P_{01}(1)$<br>$P_{12}(1) \otimes Z(1)P_{12}(1)$ | 100 001 202<br>$\xi\,1\,1\,1\,1\,\xi\,\xi^2\,1\,\xi^2$ |
| 12 | 021 111 012<br>$1\,\xi^2\,\xi\,\xi\,\xi\,\xi\,1\,\xi\,\xi^2$ | P(2)<br>W(2) | $P_{12}(1) \otimes X^T(1)$<br>$P_{12}(1) \otimes Z^*(1)$ | 100 010 220<br>$\xi\,1\,1\,1\,\xi\,1\,\xi^2\,\xi^2\,1$ |
| 13 | 012 021 111<br>$1\,\xi\,\xi^2\,1\,\xi^2\,\xi\,\xi\,\xi\,\xi$ | P(2)<br>W(2) | $I(1) \otimes P_{01}(1)$<br>$I(1) \otimes Z(1)P_{12}(1)$ | 100 202 001<br>$\xi\,1\,1\,\xi^2\,1\,\xi^2\,1\,1\,\xi$ |
| 14 | 021 012 111<br>$1\,\xi^2\,\xi\,1\,\xi\,\xi^2\,\xi\,\xi\,\xi$ | P(2)<br>W(2) | $I(1) \otimes X^T(1)$<br>$I(1) \otimes Z^*(1)$ | 100 220 010<br>$\xi\,1\,1\,\xi^2\,\xi^2\,1\,1\,\xi\,1$ |
| 15 | 012 210 222<br>$1\,\xi\,\xi^2\,\xi^2\,\xi\,1\,\xi^2\,\xi^2\,\xi^2$ | P(2)<br>W(2) | $P_{01}(1) \otimes X(1)$<br>$Z(1)P_{12}(1) \otimes Z(1)$ | 202 001 100<br>$\xi^2\,1\,\xi^2\,1\,1\,\xi\,\xi\,1\,1$ |
| 16 | 012 222 210<br>$1\,\xi\,\xi^2\,\xi^2\,\xi^2\,\xi^2\,\xi^2\,\xi\,1$ | P(2)<br>W(2) | $P_{01}(2)$<br>$Z(2)P_{12}(2)$ | 202 100 001<br>$\xi^2\,1\,\xi^2\,\xi\,1\,1\,1\,1\,\xi$ |
| 17 | 021 201 222<br>$1\,\xi^2\,\xi\,\xi^2\,1\,\xi\,\xi^2\,\xi^2\,\xi^2$ | P(2)<br>W(2) | $P_{01}(1) \otimes N(1)$<br>$Z(1)P_{12}(1) \otimes Z^*(1)P_{12}(1)$ | 220 010 100<br>$\xi^2\,\xi^2\,1\,1\,\xi\,1\,\xi\,1\,1$ |
| 18 | 021 222 201<br>$1\,\xi^2\,\xi\,\xi^2\,\xi^2\,\xi^2\,\xi^2\,1\,\xi$ | P(2)<br>W(2) | $P_{01}(1) \otimes X^T(1)$<br>$Z(1)P_{12}(1) \otimes Z^*(1)$ | 220 100 010<br>$\xi^2\,\xi^2\,1\,\xi\,1\,1\,1\,\xi\,1$ |

Class 9: Reference function $f = [\,0\,2\,0\,0\,1\,1\,0\,0\,2\,]^T = 2x_1x_2 \oplus x_2 \oplus (x_2)^2 \bmod 3$ ;

$F = [\,1\,\xi^2\,1\,1\,\xi\,\xi\,1\,1\,\xi^2\,]^T$ ; $S_f = 3\,[\,1\,1\,1\,1\,\xi\,\xi^2\,\xi^2\,\xi\,1\,]^T$; exponents of $S_f$ = [000 012 210]

| | $g$<br>$G = W(2) \cdot F$ | | | $[\![S_g = P(2)S_f]\!]$ |
|---|---|---|---|---|
| 2 | 002 011 020<br>$1\,1\,\xi^2\,1\,\xi\,\xi\,1\,\xi^2\,1$ | P(2)<br>W(2) | $I(1) \otimes P_{12}(1)$<br>$I(1) \otimes P_{12}(1)$ | 000 021 201<br>$1\,1\,1\,1\,\xi^2\,\xi\,\xi^2\,1\,\xi$ |
| 3 | 011 020 002<br>$1\,\xi\,\xi\,1\,\xi^2\,1\,1\,1\,\xi^2$ | P(2)<br>W(2) | $I(1) \otimes P_{01}(1)$<br>$I(1) \otimes Z(1)P_{12}(1)$ | 000 102 120<br>$1\,1\,1\,\xi\,1\,\xi^2\,\xi\,\xi^2\,1$ |
| 4 | 011 002 020<br>$1\,\xi\,\xi\,1\,1\,\xi^2\,1\,\xi^2\,1$ | P(2)<br>W(2) | $I(1) \otimes X^T(1)$<br>$I(1) \otimes Z^*(1)$ | 000 120 102<br>$1\,1\,1\,1\,\xi^2\,1\,\xi\,1\,\xi^2$ |
| 5 | 002 020 011<br>$1\,1\,\xi^2\,1\,\xi^2\,1\,1\,\xi\,\xi$ | P(2)<br>W(2) | $I(1) \otimes X(1)$<br>$I(1) \otimes Z(1)$ | 000 201 021<br>$1\,1\,1\,\xi^2\,1\,\xi\,1\,\xi^2\,\xi$ |
| 6 | 020 002 011<br>$1\,\xi^2\,1\,1\,1\,\xi^2\,1\,\xi\,\xi$ | P(2)<br>W(2) | $P_{12}(1) \otimes I(1)$<br>$P_{12}(1) \otimes I(1)$ | 000 210 012<br>$1\,1\,1\,\xi^2\,1\,1\,\xi\,\xi\,\xi^2$ |
| 7 | 020 110 200<br>$1\,\xi^2\,1\,\xi\,\xi\,1\,\xi^2\,1\,1$ | P(2)<br>W(2) | $P_{01}(1) \otimes I(1)$<br>$Z(1)P_{12}(1) \otimes I(1)$ | 012 000 210<br>$1\,\xi\,\xi^2\,1\,1\,1\,\xi^2\,\xi\,1$ |
| 8 | 020 200 110<br>$1\,\xi^2\,1\,\xi^2\,1\,1\,\xi\,\xi\,1$ | P(2)<br>W(2) | $X^T(1) \otimes I(1)$<br>$Z^*(1) \otimes I(1)$ | 012 210 000<br>$1\,\xi\,\xi^2\,\xi^2\,\xi\,1\,1\,1\,1$ |
| 9 | 002 101 200<br>$1\,1\,\xi^2\,1\,\xi\,\xi^2\,1\,1$ | P(2)<br>W(2) | $P_{01}(1) \otimes P_{12}(1)$<br>$Z(1)P_{12}(1) \otimes P_{12}(1)$ | 021 000 201<br>$1\,\xi^2\,\xi\,1\,1\,1\,\xi^2\,1\,\xi$ |
| 10 | 002 200 101<br>$1\,1\,\xi^2\,\xi^2\,1\,1\,\xi\,1\,\xi$ | P(2)<br>W(2) | $I(1) \otimes X(1)$<br>$I(1) \otimes Z(1)$ | 021 201 000<br>$1\,\xi^2\,\xi\,\xi^2\,1\,\xi\,1\,1\,1$ |
| 11 | 011 110 212<br>$1\,\xi\,\xi\,\xi\,\xi\,1\,\xi^2\,\xi\,\xi^2$ | P(2)<br>W(2) | $P_{01}(2)$<br>$Z(2)P_{12}(2)$ | 102 000 120<br>$\xi\,1\,\xi^2\,1\,1\,1\,\xi\,\xi^2\,1$ |
| 12 | 011 212 110<br>$1\,\xi\,\xi\,\xi^2\,\xi\,\xi^2\,\xi\,\xi\,1$ | P(2)<br>W(2) | $N(1) \otimes X^T(1)$<br>$Z^*(1)P_{12}(1) \otimes Z^*(1)$ | 102 120 000<br>$\xi\,1\,\xi^2\,\xi\,\xi^2\,1\,1\,1\,1$ |
| 13 | 011 101 221<br>$1\,\xi\,\xi\,\xi\,1\,\xi\,\xi^2\,\xi^2\,\xi$ | P(2)<br>W(2) | $P_{01}(1) \otimes X^T(1)$<br>$Z(1)P_{12}(1) \otimes Z^*(1)$ | 120 000 102<br>$\xi\,\xi^2\,1\,1\,1\,1\,\xi\,1\,\xi^2$ |
| 14 | 011 221 101<br>$1\,\xi\,\xi\,\xi^2\,\xi^2\,\xi\,\xi\,1\,\xi$ | P(2)<br>W(2) | $N(1) \otimes P_{01}(1)$<br>$Z^*(1)P_{12}(1) \otimes Z(1)P_{12}(1)$ | 120 102 000<br>$\xi\,\xi^2\,1\,\xi\,1\,\xi^2\,1\,1\,1$ |
| 15 | 002 122 212<br>$1\,1\,\xi^2\,\xi\,\xi^2\,\xi^2\,\xi^2\,\xi\,\xi^2$ | P(2)<br>W(2) | $P_{01}(1) \otimes X(1)$<br>$Z(1)P_{12}(1) \otimes Z(1)$ | 201 000 021<br>$\xi^2\,1\,\xi\,1\,1\,1\,1\,\xi^2\,\xi$ |
| 16 | 002 212 122<br>$1\,1\,\xi^2\,\xi^2\,\xi\,\xi^2\,\xi\,\xi^2\,\xi^2$ | P(2)<br>W(2) | $N(1) \otimes P_{12}(1)$<br>$Z^*(1)P_{12}(1) \otimes P_{12}(1)$ | 201 021 000<br>$\xi^2\,1\,\xi\,1\,\xi^2\,\xi\,1\,1\,1$ |
| 17 | 020 122 221<br>$1\,\xi^2\,1\,\xi\,\xi^2\,\xi^2\,\xi^2\,\xi^2\,\xi$ | P(2)<br>W(2) | $X(1) \otimes I(1)$<br>$Z(1) \otimes I(1)$ | 210 000 012<br>$\xi^2\,\xi\,1\,1\,1\,1\,1\,\xi\,\xi^2$ |
| 18 | 020 221 122<br>$1\,\xi^2\,1\,\xi^2\,\xi^2\,\xi\,\xi\,\xi^2\,\xi^2$ | P(2)<br>W(2) | $N(1) \otimes I(1)$<br>$Z^*(1)P_{12}(1) \otimes I(1)$ | 210 012 000<br>$\xi^2\,\xi\,1\,1\,\xi\,\xi^2\,1\,1\,1$ |